\documentclass[letterpaper,preprintnumbers,superscriptaddress,aps,nofootinbib,perprint,12pt]{revtex4}
\usepackage{CJK,upgreek,fancyhdr}
\usepackage{amssymb}
\usepackage[centertags]{amsmath}
\usepackage{txfonts}
\usepackage{epsfig}
\usepackage{cases}
\usepackage{slashed}
\usepackage{amsfonts}

\usepackage{bm}
\usepackage{color}
\usepackage{graphicx,graphics}
\usepackage{multirow}
\usepackage{float}
\usepackage[pdfstartview=FitH]{hyperref}
\usepackage{slashed}
\usepackage{ulem}
\usepackage{booktabs}
\usepackage{multirow}
\usepackage{dcolumn}
\usepackage{subfig}
\usepackage{comment}

\hypersetup{colorlinks=true, citecolor=blue, linkcolor=blue,filecolor=black,urlcolor=blue}

\def\met{\displaystyle{\slash}\!\!\! \!E_T}

\begin{document}
\begin{CJK*}{GB}{gbsn}

\title{Search for $W^{\prime}$ signal in single top quark production at the LHC }

\author{Fei Huang (»Æ·É)}\affiliation{School of Physics, Shandong University, Jinan Shandong 250100,  China}
\author{Hong-Lei Li (ÀîºéÀÙ)}\affiliation{School of Physics and Technology, University of Jinan, Jinan Shandong 250022,  China}
\author{Shi-Yuan Li (ÀîÊÀÔ¨)}\affiliation{School of Physics, Shandong University, Jinan Shandong 250100,  China}
\author{Zong-Guo Si (˾×Ú¹ú)}\affiliation{School of Physics, Shandong University, Jinan Shandong 250100,  China}\affiliation{CAS Key Laboratory of Theoretical Physics, Institute of Theoretical Physics,
Chinese Academy of Sciences, Beijing 100190, China}
\author{Wei Su (ËÕΰ)}\affiliation{CAS Key Laboratory of Theoretical Physics, Institute of Theoretical Physics,
Chinese Academy of Sciences, Beijing 100190, China}\affiliation{School of Physics, University of Chinese Academy of Sciences, Beijing 100049, China}
\author{Zhong-Juan Yang (ÑîÖоê)}\affiliation{School of Physics and Technology, University of Jinan, Jinan Shandong 250022,  China}
\begin{abstract}
Heavy charged gauge bosons are proposed in some theories beyond the standard model. We explore the discovery potential for $W'\to t\bar{b}$ with top quark semi-leptonic decay at the LHC. We concentrate on the new physics signal search with the deviation from the standard model prediction if the resonance peak of $W'$ cannot be observed directly. Signal events  with two jets plus one charged lepton and missing energy are simulated, together with the dominant standard model backgrounds. In this paper, it is found that suitable cuts on the kinematic observables can effectively suppress the standard model backgrounds, so that it is possible to search for a $W'$ signal at the LHC if its mass is less than 6.6 TeV.
\end{abstract}


\maketitle
\section{INTRODUCTION}\label{sec1}
 Gauge sector extension is one of the promising new physics theories beyond the standard model (SM). Heavy charged gauge bosons ($W^{\prime \pm}$) are involved in a number of the new physics models, such as Extra Dimensions~\cite{Klein:1926tv,ArkaniHamed:1998rs,Randall:1999vf,Randall:1999ee,ArkaniHamed:2001ca,Appelquist:2000nn,Cheng:2002ab}, Little Higgs~\cite{ArkaniHamed:2001nc,Han:2003wu,Kaplan:2003uc}, GUTs~\cite{Pati:1973uk,Georgi:1974sy,Fritzsch:1974nn}, etc. A simple but well-motivated scenario is the Left-Right symmetric model~\cite{Pati:1974yy,Mohapatra:1974hk,Mohapatra:1974gc,Senjanovic:1975rk,Mohapatra:1977mj}, which is based on the extended $SU(2)_L\times SU(2)_R \times U(1)$ gauge group.
 Provided the current experimental constraints, a TeV-scaled charged gauge boson is allowed, which provides the opportunity for new physics searches at the LHC.

 The leptonic decay $ W' \to l \nu$ is the golden channel for searching for $W'$ if the couplings to the SM leptons are not specifically suppressed. According to a $M_T$-distribution, determined by the transverse momentum of the charged leptons and missing transverse energy, lower mass limits of 5.1 (4.1) TeV for the sequential SM $W'$ boson have been obtained by the ATLAS and CMS collaborations at $\sqrt{s} =13$ TeV LHC~\cite{Aaboud:2017efa,Khachatryan:2016jww}. Although the leptonic decay modes are experimentally clean and possibly may be the first observed, the other decay channels need to be studied in depth to understand the properties of the heavy bosons, especially in some leptonic branch ratio suppressed scenarios. Although the light quark decay modes of $W' \to q \bar q'$ have a larger production rate than the $W' \to t \bar b$ channel, there is no advantage for searches for the $W'$ boson due to the large QCD backgrounds at the LHC. Furthermore,  the $W' \to t \bar b$ mode has a characteristic jet-substructure with the top quark, and a large number of events with single top quark production can be accumulated at the LHC~\cite{Sullivan:2013ina,Duffty:2012rf}.

 If the $W'$ is discovered at the LHC, it becomes imperative to investigate the details of its intrinsic properties and its interactions with other particles. The chiral couplings to standard model fermions are  crucial features which differ from the SM weak interactions in some specific models. It has been demonstrated that the angular distributions of the top quark and lepton resulting from top decay can be used to disentangle the chiral couplings of the $W'$ to SM fermions with the $W' \to t \bar b$ mode~\cite{Gopalakrishna:2010xm}. We have also found that the charged lepton angular distribution can be used to distinguish the chirality of $W'$ in the decay mode of $W'\to WH \to b \bar b l \nu$~\cite{Bao:2011nh}. The investigation of the $W'$ boson has also been extended to the associated production or exotic decay modes~\cite{Gopalakrishna:2010xm,Gong:2014qla,Berger:2011hn,Berger:2011xk,Kelso:2014qka,Ayazi:2010jd}.

 Recently, the CMS collaboration has reported the latest results on the search for a resonance peak with $W' \to t \bar b$~\cite{Sirunyan:2017ukk}. The right-handed $W'$ boson is excluded for mass less than 2.6 TeV with  the top quark decaying hadronically and leptonically. Unfortunately, no evidence of the $W'$ resonance peak can be observed directly up to now.  Motivated by the reach of the $W'$ investigation at the LHC, we provide various strategies to search for a significant excess from the standard model prediction in kinematics distributions other than the new resonance peak. We propose four schemes based on different cuts to suppress the standard model backgrounds. Cuts on the transverse momentum of jets ($p_{T}^j$), invariant mass of jets ($M_{jj}$), collision energy scale ($H_T$) and invariant mass of top and bottom quark ($M_{t\bar{b}}$) are adopted to highlight the signal process. We find that the lower mass limit for the sequential $W'$ boson is up to 3.7-6.6 TeV.

 This paper is arranged as follows. In Section \ref{sec2} we briefly depict the theoretical framework and show the difference between the $W'_L$ and $W'_R$ bosons. The detector simulation and numerical results with various schemes are presented in Section~\ref{sec3}. Finally, a short summary  is given in Section ~\ref{sec4}.
\section{THEORETICAL FRAMEWORK}\label{sec2}
 Heavy charged gauge bosons are predicted in many new physics theories. Provided that the SM is an approximation of the new physics in the low energy scale,  the most direct detection for new physics should be via the decay of these heavy particles into the SM particles. The relevant gauge interactions between  $W'$ and fermions can be generalized in the formula
  \begin{align}
  \mathcal{L}=g_{L}\frac{g_2}{\sqrt{2}}\bar{\psi_{u}^{i}}\gamma_{\mu}V_{L}^{\prime ij}\frac{1}{2}(1-\gamma_{5})\psi_{j}^{d}W_{L}^{\prime}+g_{R}\frac{ g_2}{\sqrt{2}}\bar{\psi_{u}^{i}}\gamma_{\mu}V_{R}^{\prime ij}\frac{1}{2}(1+\gamma_{5})\psi_{j}^{d}W_{R}^{\prime}+H.c. , \label{f:Lagrangian}
  \end{align}
 where $g_2$ is the SM electroweak coupling and $g_{L}$ ($g_{R}$) is the left-handed (right-handed) coupling constant, with $g_{L}=1$, $g_{R}=0$ the pure left-handed gauge interaction (labeled $W_{L}^{\prime}$ ) and $g_{L}=0$, $g_{R}=1$ the pure right-handed gauge interaction (labeled   $W_{R}^{\prime}$).  $V^{\prime}$ is the flavor mixing matrix, the counterpart of the Cabibbo-Kobayashi-Maskawa matrix in the SM.

 Both left- and right-handed $W'$ bosons can exist in the left-right symmetric model, as well as the right-handed fermion doublets, which lead to a heavy neutrino ($N$). As discussed in Ref.~\cite{Han:2012vk}, if the $W'$ is heavier than $N$, the decay mode of $W' \to l N$ is open, which provides an interesting like-sign dilepton production process to learn the lepton number violation. Otherwise, we can only investigate the $W'$ boson from its couplings to SM particles, with the $W' \to l N$ decay modes forbidden. Thus the three dominant  decay modes  are $W^{\prime}\rightarrow t\bar{b}$, $W^{\prime}\rightarrow q\bar{q}'$, and $W^{\prime}\rightarrow \ell\nu$. The right-handed $W'$ has the same decay modes as the left-handed one except for $W^{\prime}\rightarrow \ell\nu$, since the right-handed neutrino is absent in the SM. The $W^{\prime}_{L}$ has a larger decay width than $W^{\prime}_{R}$, which is expressed in the following formulae
 \begin{align}
 \Gamma_{W_{R}^{\prime}}=\frac{g_2^2g_R^2m_{W^{\prime}}}{16\pi}\Bigr[2+\Big(1-\frac{m_t^2}{m_{W^{\prime}}^2}\Big)
 \Big(1-\frac{m_t^2}{2m_{W^{\prime}}^2}-\frac{m_t^4}{2m_{W^{\prime}}^4}\Big)\Bigr], 
 \nonumber \\
 \Gamma_{W_{L}^{\prime}}=\frac{g_2^2g_L^2m_{W^{\prime}}}{16\pi}\Bigr[3+\Big(1-\frac{m_t^2}{m_{W^{\prime}}^2}\Big)
 \Big(1-\frac{m_t^2}{2m_{W^{\prime}}^2}-\frac{m_t^4}{2m_{W^{\prime}}^4}\Big)\Bigr], \label{f:GammaWLprime}
 \end{align}
 where $m_W'$ ($m_t$) is the mass of $W'$ boson (top quark).

 In this paper, we focus on the process
  \begin{equation}
 pp\rightarrow W^{\prime +} \slash  W^{+} \rightarrow \bar{b} t \rightarrow  \bar{b} b l^{+}\nu, ~~~~l^+=e^+, \mu^+. \label{eq:process}
 \end{equation}
 \begin{figure}[ht!]
  \centering
  \includegraphics[width=10cm]{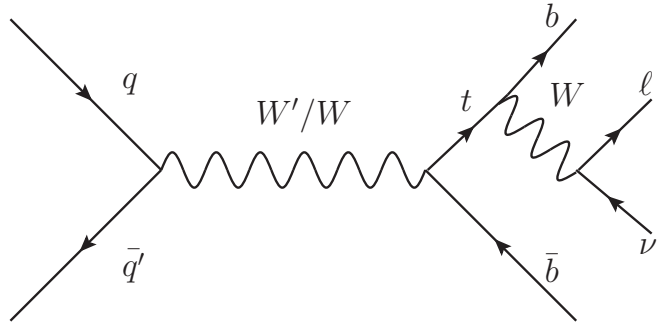}\\
  \caption{Feynman diagram of process (\ref{eq:qprocess}).}\label{fig:wp1}
\end{figure}

\begin{figure}
    \centering
   \subfloat[]{
   \begin{minipage}[t]{0.4\textwidth}
   \centering
    \includegraphics[width=7cm,height=6cm]{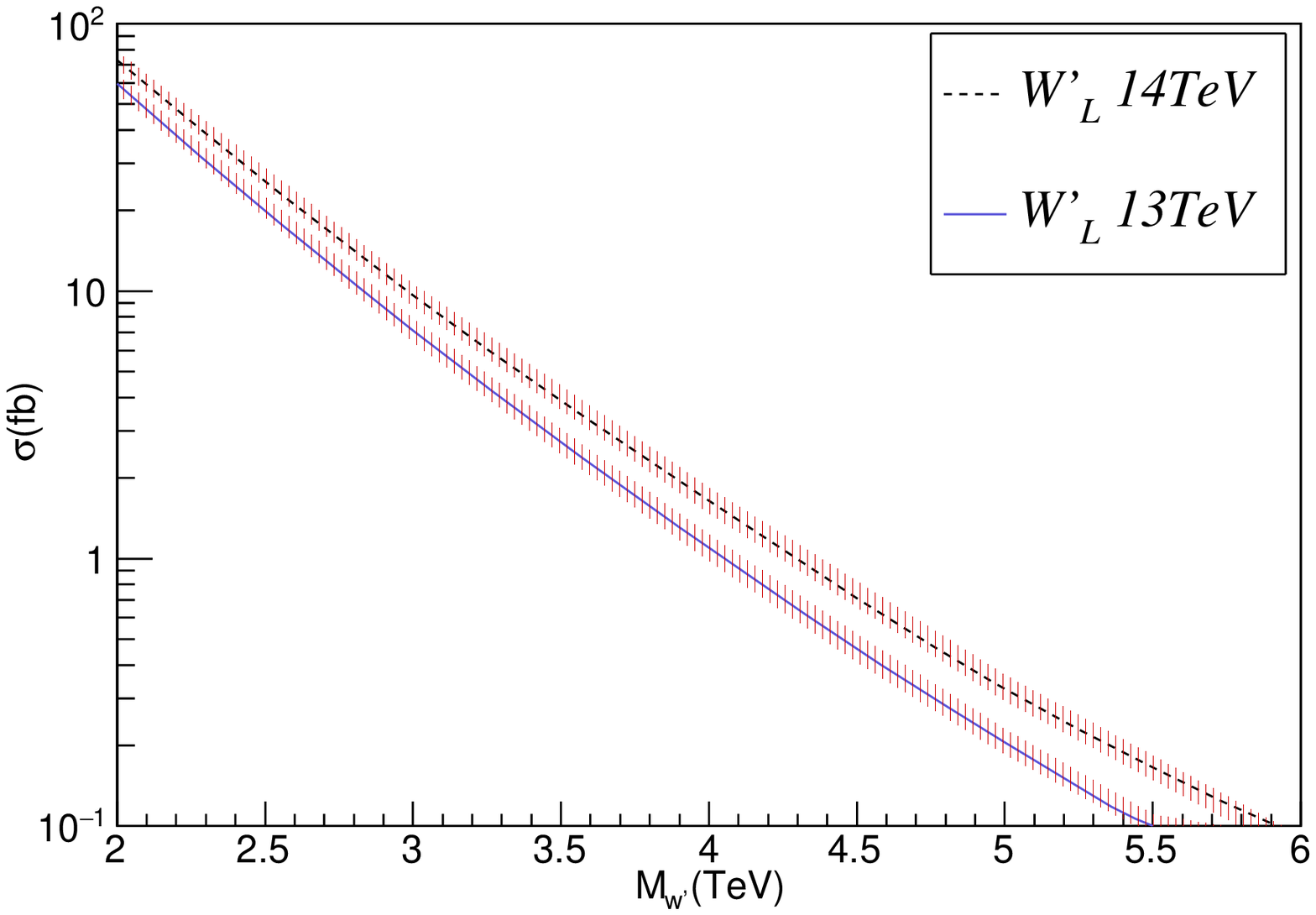}
   \end{minipage}
   }
   \subfloat[]{
   \begin{minipage}[t]{0.4\textwidth}
   \centering
   \includegraphics[width=7cm,height=6cm]{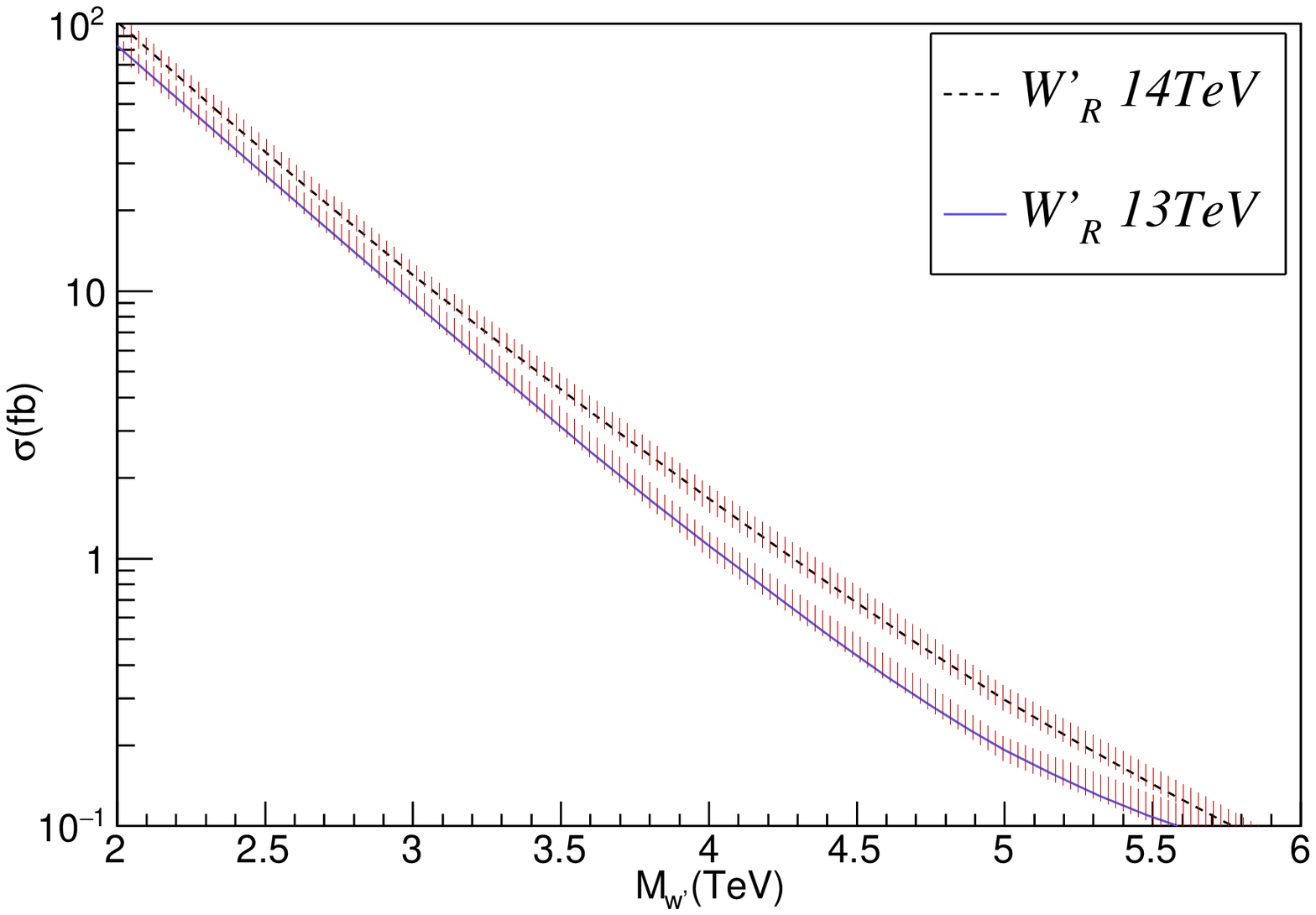}
   \end{minipage}
   }
   \caption{The cross section  of process  $pp\rightarrow W^{\prime +} \rightarrow \bar{b} t \rightarrow  \bar{b} b l^{+}\nu ~(l^+=e^+, \mu^+) $ with respect to the $W'$ mass at the LHC. The shaded region represents the uncertainty from the PDF with the energy scale varying from $\sqrt{S}/2$ to $2\sqrt{S}$. (a) $W'=W'_{L}$ without the effects of $W$; (b)  $W'=W'_{R}$.}\label{fig:onlysignal}
  \end{figure}

 The corresponding total cross section can be written as
 \begin{align}
 \sigma=\int f_{q}(x_{1})f_{\bar{q^{\prime}}}(x_2)\hat{\sigma}(\sqrt{x_1x_2S}~)dx_{1}dx_{2},
 \end{align}
 where $f_{q \slash {\bar{q}'}}(x_{i})$ is the parton distribution function (PDF) with $x_i$ the parton momentum fraction.  $\sqrt{S}$ is the proton-proton collision center of mass energy. $\hat{\sigma}$ represents the partonic cross section of the process
\begin{equation}
 q(p_1) + \bar{q'}(p_2) \rightarrow W^{\prime +} \slash  W^{+} \rightarrow \bar{b}(p_3) + t(p_t) \rightarrow  \bar{b}(p_3) + b(p_4) + l^{+}(p_5) + \nu(p_6), \label{eq:qprocess}
 \end{equation}
 where $p_i$(i=1,2,3,4,5,6) is the momentum of the corresponding particle, and $p_t$ is the momentum of the top quark. The corresponding Feynman diagram is shown in Fig.~{\ref{fig:wp1}} with the differential cross section
 \begin{equation}
 d\hat{\sigma}=\frac{1}{2s}\overline{|\mathcal{M}|^2}d \mathcal{L}ips_{4},
 \end{equation}
 where $s=x_1x_2S$, and $\mathcal{L}ips_{4}$ denotes the Lorentz invariant phase space of the four final particles. $\overline{|\mathcal{M}|^2}$ represents the invariant amplitude of the partonic process~(\ref{eq:qprocess}) summed (averaged) over the final (initial) particle colors and spins, and can be written as,
\begin{numcases}{\overline{{|\mathcal{M}|^2}}=}
\overline{|\mathcal{M}_{W'_L}|^2} + \overline{|\mathcal{M}_{W}|^2} + 2Re(\mathcal{M}_{W'_L}^{*}\mathcal{M}_W),&for $W'_{L}$;\notag \\
\overline{|\mathcal{M}_{W'_R}|^2} + \overline{|\mathcal{M}_{W}|^2},&for $W'_{R}$,
\end{numcases}
 where  $\overline{{|\mathcal{M}_i}|^2}$ ($i= W'_L, W'_R, W$) is the corresponding invariant amplitude and $2Re(\mathcal{M}_{W'_L}^{*}\mathcal{M}_W)$ is the interference term between $W'_L$ and $W$,
\begin{align}
&\overline{|\mathcal{M}_{W'_L}|^2 }=
  \frac{32g_{L}^4g_2^8(p_1\cdot p_3)(p_4\cdot p_5) \big(2(p_2\cdot p_{t})(p_{t}\cdot p_6)-p_{t}^2(p_2\cdot p_6\big))}{[(p_{t}^2-m_t^2)^2+m_t^2\Gamma_{t}^2][(p_{W}^2-m_W^2)^2+m_W^2\Gamma_{W}^2]
  [(s-m_{W^{\prime}}^2)^2+m_{W^{\prime}}^2\Gamma_{W^{\prime}}^2]},  \nonumber \\
&\overline{|\mathcal{M}_{W'_R}|^2} =\frac{32g_{R}^4m_t^2g_2^8(p_1\cdot p_3)(p_2\cdot p_6)(p_4\cdot p_5)}{[(p_{t}^2-m_t^2)^2+m_t^2\Gamma_{t}^2][(p_{W}^2-m_W^2)^2+m_W^2\Gamma_{W}^2][(s-m_{W^{\prime}}^2)^2+m_{W^{\prime}}^2\Gamma_{W^{\prime}}^2]}, \nonumber \\
&\overline{|\mathcal{M}_{W}|^2} =\frac{32g_2^8(p_1\cdot p_3)(p_4\cdot p_5) \big(2(p_2\cdot p_{t})(p_{t}\cdot p_6)-p_{t}^2(p_2\cdot p_6\big))}{[(p_{t}^2-m_t^2)^2+m_t^2\Gamma_{t}^2][(p_{W}^2-m_W^2)^2+m_W^2\Gamma_{W}^2]
  [(s-m_{W}^2)^2+m_{W}^2\Gamma_{W}^2]}, \nonumber \\
&2Re(\mathcal{M}_{W'_L}^{*}\mathcal{M}_W)
 =\Bigg\{\frac{2\big[(s-m_W^2)(s-m_{W^{\prime}}^2)+m_W\Gamma_Wm_{W^{\prime}}\Gamma_{W^{\prime}}\big]}
{[(s-m_{W^{\prime}}^2)^2+M_{W^{\prime}}^2\Gamma_{W^{\prime}}^2]
    [(s-m_{W}^2)^2+m_{W}^2\Gamma_{W}^2]} \Bigg\} \notag\\
    &~~~~~~~~~~~~~~~~~~~~~~~~~~~~\frac{32g_{L}^2g_2^8(p_1\cdot p_3)(p_4\cdot p_5) \big(2(p_2\cdot p_{t})(p_{t}\cdot p_6)-p_{t}^2(p_2\cdot p_6\big))}{[(p_{t}^2-m_t^2)^2+m_t^2\Gamma_{t}^2][(p_{W}^2-m_W^2)^2+m_W^2\Gamma_{W}^2]}.
\end{align}

 The couplings of $g_{L(R)}$ are arbitrary in various models, while  the Sequential $W'$ model with the $W'$ boson has the same couplings to quarks and leptons as the $W$ boson. We have numerical results in the framework of the Sequential $W'$ model. CTEQ6L1~\cite{Pumplin:2002vw} is set for PDF, with $m_{W}=80.4$ GeV and $m_t=173.1$ GeV~\cite{Patrignani:2016xqp}. The cross section  of the process  $pp\rightarrow W^{\prime +} \rightarrow \bar{b} t \rightarrow  \bar{b} b l^{+}\nu ~(l^+=e^+, \mu^+)$ with respect to the $W'$ mass at 13 and 14 TeV  is shown in Fig.~\ref{fig:onlysignal}. There are more than ten events produced with a $W'$ mass around 6 TeV with a luminosity of 300 $fb^{-1}$. The shaded region represents the uncertainty from the PDF with the energy scale varying from $\sqrt{S}/2$ to $2\sqrt{S}$. This uncertainty could affect the cross section by about 10$\sim$20\% with the tree level result. It will decrease with the higher order calculation, which is out of the scope of this work. So in the following work we focus on the investigation of $W'$ at 14 TeV and assume a luminosity of 300 $fb^{-1}$ unless otherwise stated.
\section{NUMERICAL RESULTS AND DISCUSSION}\label{sec3}
 Once the $W'$ boson is produced at the LHC, the $W' \to t \bar{b}$ channel will play an important role in the search for $W'$ signal in the large $W'$ mass region. In this work, we provide various strategies to investigate the lower limit on the $W'$ mass from $t \bar{b}$ production with the signal of $2~\text{jets}+1~\text{lepton}+~\met$.

\begin{figure}
    \centering
   \subfloat[]{
   \begin{minipage}[t]{0.4\textwidth}
   \centering
    \includegraphics[width=7cm,height=6cm]{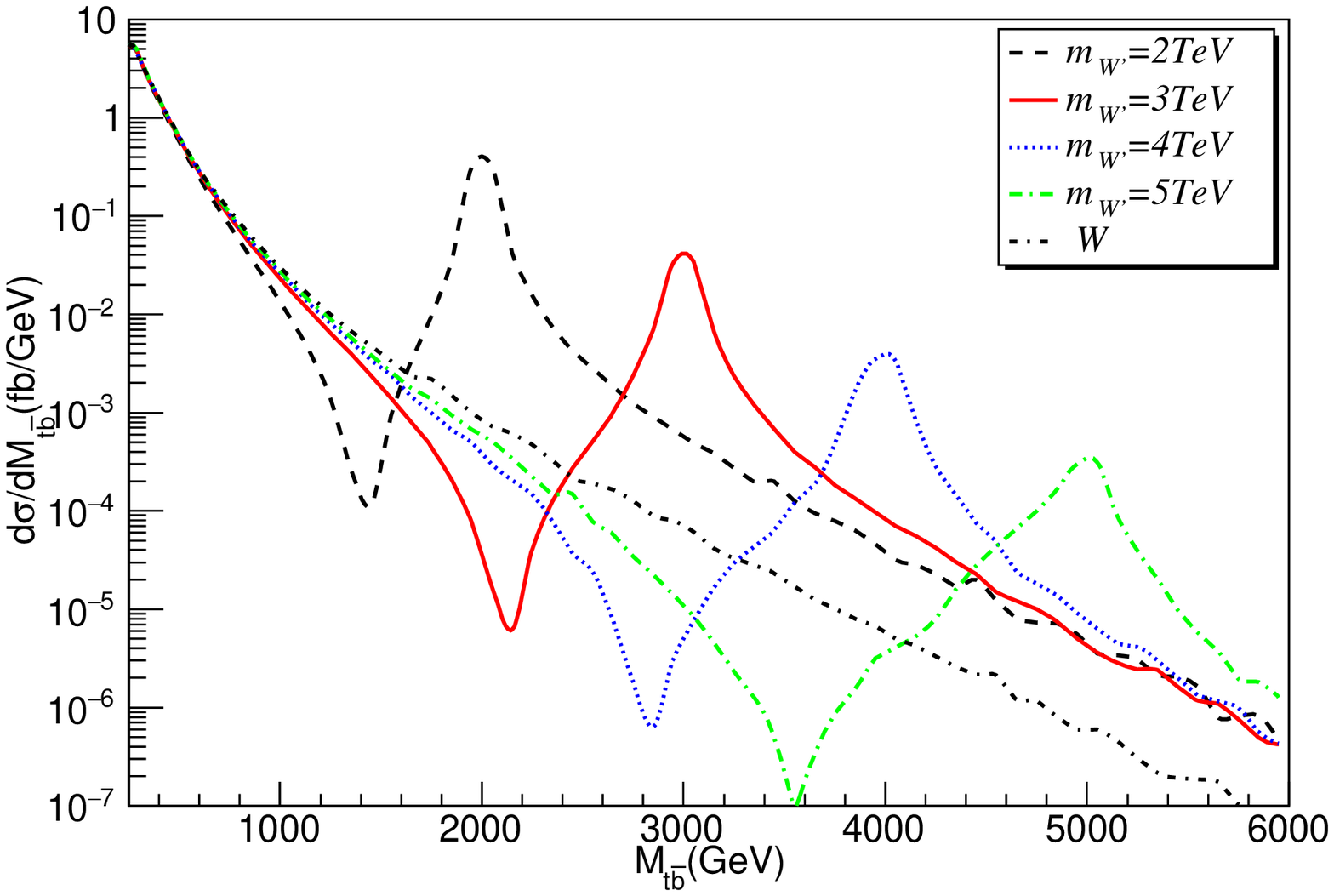}
   \end{minipage}
   }
   \subfloat[]{
   \begin{minipage}[t]{0.4\textwidth}
   \centering
   \includegraphics[width=7cm,height=6cm]{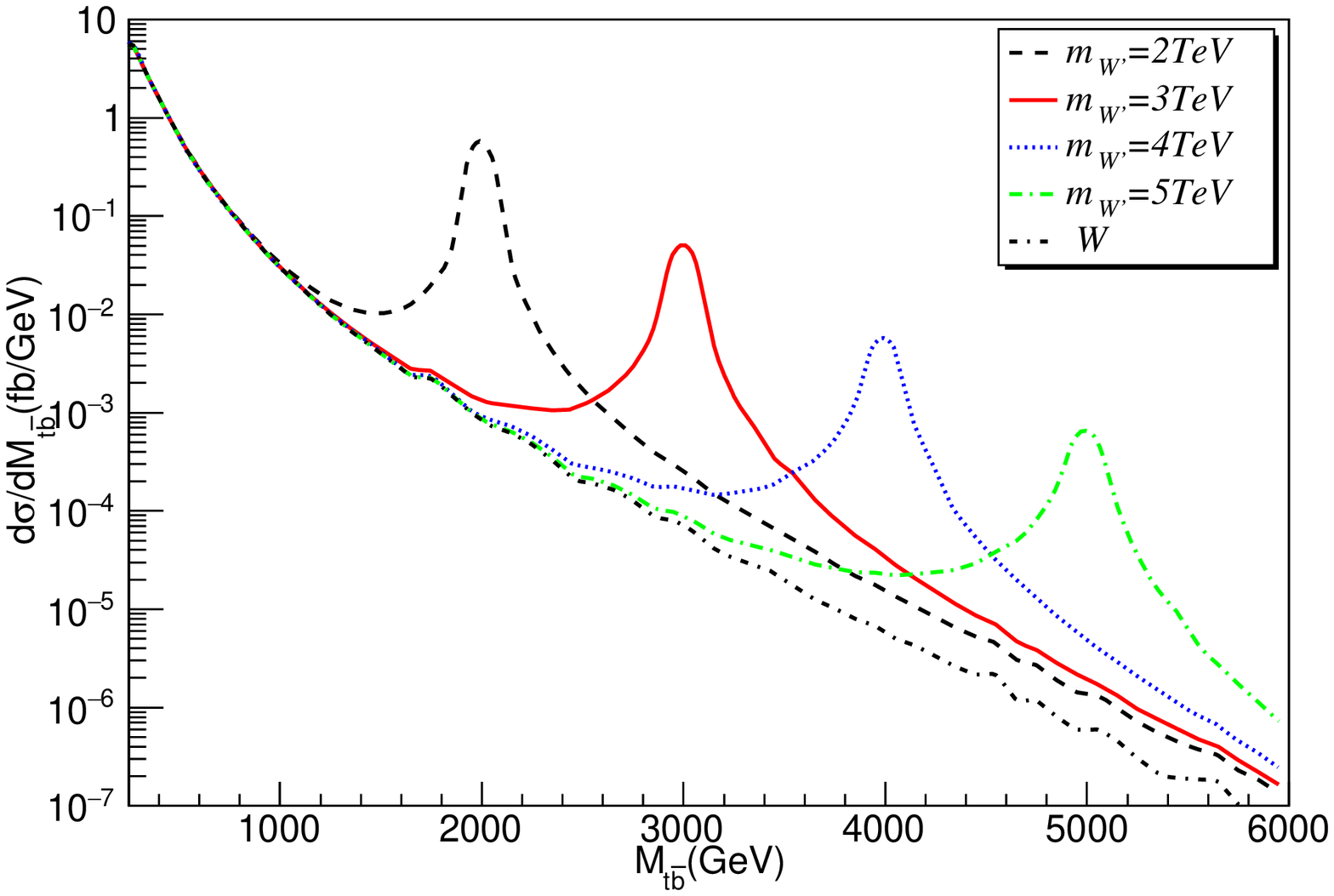}
   \end{minipage}
   }
   \caption{The invariant mass $M_{t\bar{b}}$-distribution with $m_{W^{\prime}}= 2, 3, 4, 5$ TeV at 14 TeV for the process $ pp\rightarrow W^{\prime +} \slash  W^{+} \rightarrow \bar{b} t \rightarrow  \bar{b} b l^{+}\nu, ~~~~l^+=e^+, \mu^+$. (a) $W'=W'_{L}$; (b) $W'=W'_{R}$. }\label{fig:nocutleft}
  \end{figure}

 The resonance peak through the invariant mass of $M_{t\bar{b}}$ can be reconstructed as shown in Fig.~\ref{fig:nocutleft}. The differential distributions with the invariant mass of $M_{t\bar{b}}$ between the $W'_{L}+W$ and $W'_{R}+W$ differ from the interference term. The valley region is due to the negative contribution from the interference term in the mass region of $m_W<M_{t\bar{b}}<m_{W'}$ for $W'_L$, whereas there is no interference term between $W'_R$ and the $W$ boson. This kind of phenomena can be used to distinguish $W'_{L}$ from $W'_{R}$ if enough events are accumulated. Moreover, there are a large number of SM $W$ bosons in the $t\bar{b}$ production compared with $W'$, especially in the small $M_{t\bar{b}}$ region. It is therefore crucial to suppress the influence of  $W$ bosons in the search for the $W'$ boson.
\begin{figure}[ht!]
  \centering
  \includegraphics[width=10cm]{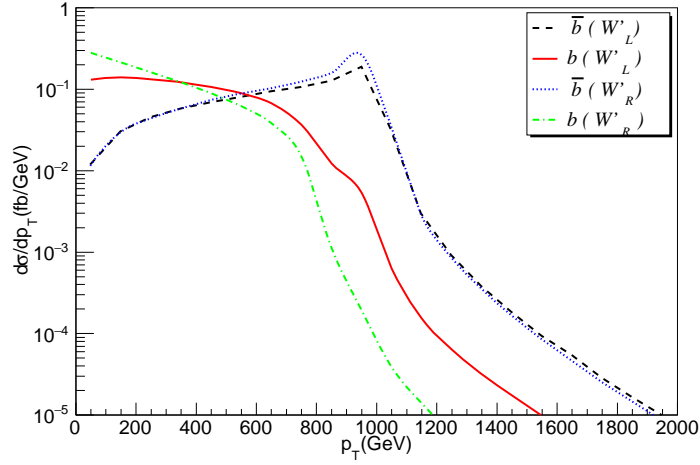}\\
  \caption{The $p_T$-distribution of b and $\bar{b}$ quark related to the process $pp\rightarrow W^{\prime +} \rightarrow \bar{b} t \rightarrow  \bar{b} b l^{+}\nu ~(l^+=e^+, \mu^+) $ with $m_{W^{\prime}}=2$ TeV at 14 TeV.}\label{fig:bbar}
\end{figure}

 Figure~{\ref{fig:bbar}} shows the transverse momentum ($p_T$) distribution of the $\bar{b}$ and b quarks  related to the process $pp\rightarrow W^{\prime +} \rightarrow \bar{b} t \rightarrow  \bar{b} b l^{+}\nu ~(l^+=e^+, \mu^+) $ with $m_{W'}=2$ TeV. The $\bar{b}$ quark distribution has a peak around 1 TeV, since for a parent particle of mass $M$ decaying to two light particles, there is a Jacobian peak near $M/2$ in the transverse momentum distribution of final state particles. Such distributions can be used to set cuts to suppress the backgrounds. In addition, the b quark distribution shows differences between $W'_L$ and $W'_R$ because of the top quark spin correlation effects, which provides the opportunity to distinguish the chirality of the $W'$ boson~\cite{Gopalakrishna:2010xm}.

 To be as realistic as possible, we simulate the detector performance by smearing the lepton and jet energies based on the assumption of Gaussian resolution parametrization
 \begin{align}
 \frac{\delta(E)}{E}=\frac{a}{\sqrt{E}}\oplus b
 \end{align}
 where $\delta(E)/E$ is the energy resolution, $a$ is a sample term, $b$ is a constant term, and $\oplus$ denotes a sum in quadrature. We always use $a=5\%, b=0.55\%$ for leptons and $a=100\%, b=5\%$ for jets~\cite{Aad:2009wy}.
 In order to identify an isolated jet or lepton, we define the angular separation between particle i and particle j as
 \begin{align}
 \triangle R_{ij}=\sqrt{\triangle \phi_{ij}^{2}+\triangle \eta_{ij}^{2}},
 \end{align}
 where $\triangle \phi_{ij}$ and $\eta_{ij}$ are the difference in azimuthal angle and rapidity between the related particles.

 For the process in Eq.~(\ref{eq:qprocess}), $W^{\prime}$ decays to two particles which are back to back in the transverse plane. The $W$ boson and bottom quark are collimated, because the top quark is highly boosted, so  the angular separation $\triangle R_{\ell b}$ between the charged lepton and bottom quark is peaked at a low value and the angular separation $\triangle R_{bb}$ between the bottom quark and bottom anti-quark is peaked near $\pi$. Therefore, we impose the basic cuts as
 \begin{align}
 \triangle R_{\ell b} > 0.3, ~   \triangle R_{bb} > 0.4, ~P_{T}^{\ell} > 20 ~\text{GeV},  \nonumber \\
 ~P_{T}^j > 50 ~\text{GeV},  ~ \eta(j) < 3.0,  ~  \met > 25 ~\text{GeV}.
 \end{align}
\begin{figure}
    \centering
   \subfloat[]{
    \label{fig:basiccutL}
   \begin{minipage}[t]{0.4\textwidth}
   \centering
    \includegraphics[width=7cm,height=6cm]{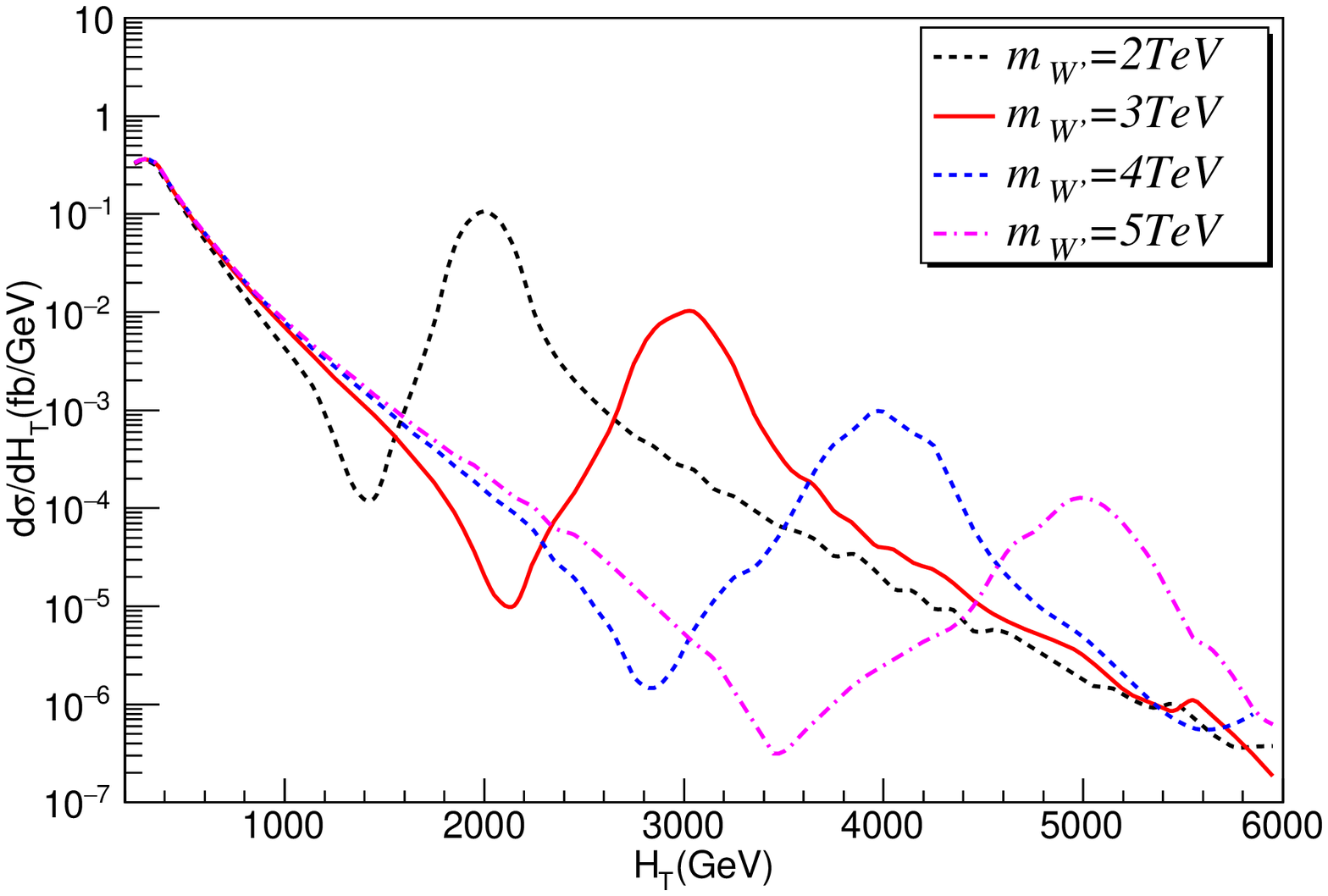}
   \end{minipage}
   }
   \subfloat[]{
   \label{fig:basiccutR}
   \begin{minipage}[t]{0.4\textwidth}
   \centering
   \includegraphics[width=7cm,height=6cm]{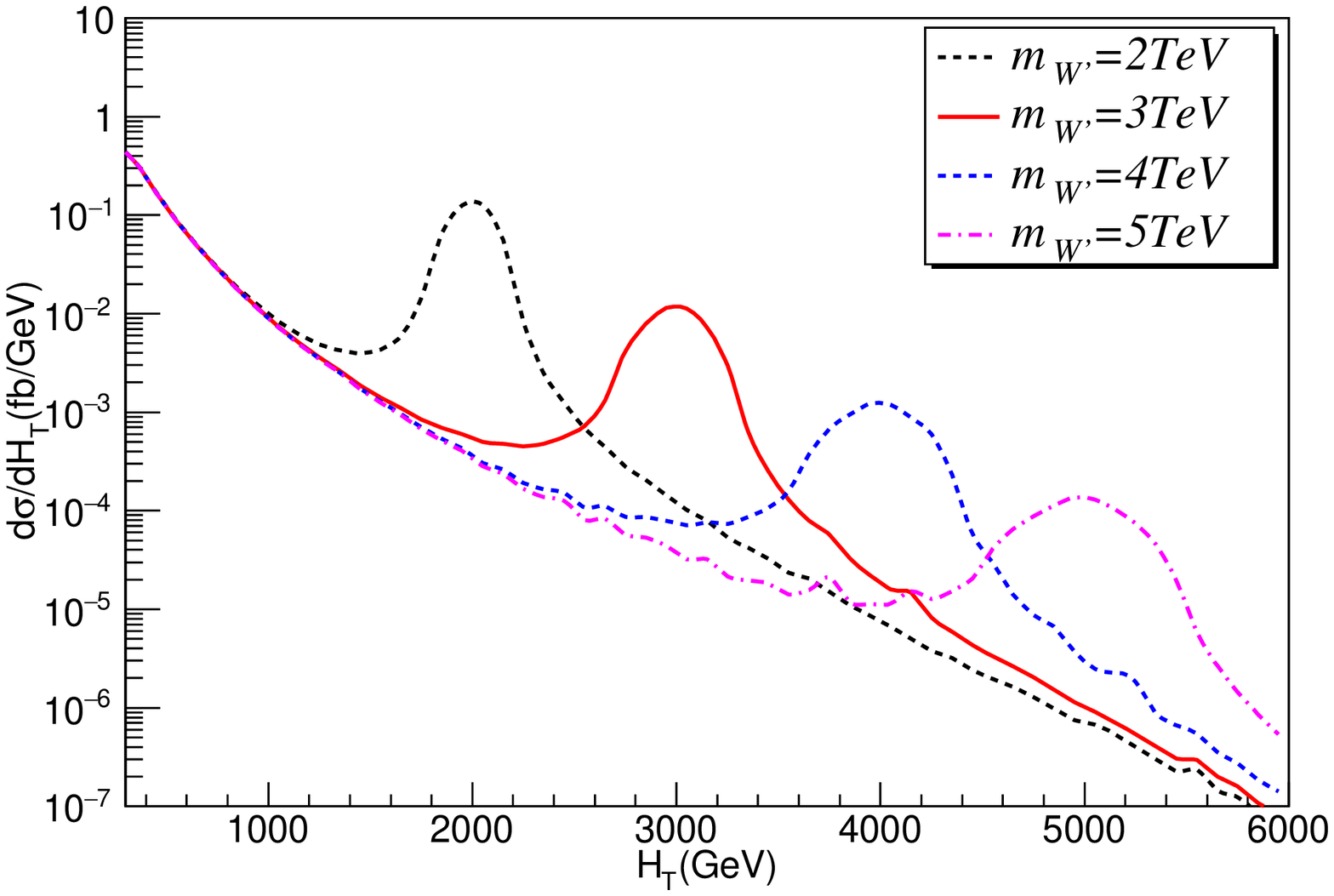}
   \end{minipage}
   }
   \caption{The invariant mass $M_{t\bar{b}}$-distribution for $m_{W^{\prime}}= 2, 3, 4, 5$ TeV at 14 TeV with the basic cuts for the process $ pp\rightarrow W^{\prime +} \slash  W^{+} \rightarrow \bar{b} t \rightarrow  \bar{b} b l^{+}\nu, ~~~~l^+=e^+, \mu^+$. (a) $W'=W'_{L}$; (b) $W'=W'_{R}$.}\label{fig:basiccutL}
  \end{figure}
 Figure~{\ref{fig:basiccutL}} shows the invariant mass ($M_{t\bar{b}}$) distribution for $m_{W^{\prime}}= 2, 3, 4, 5$ TeV at 14 TeV with the basic cuts. Compared with Fig.~{\ref{fig:nocutleft}}, the discrepancy of the peak between $W'$ and $W$ boson is weakened after the basic cuts due to more events with small transverse momentum being generated in the $W$ boson process.

 Besides the $W$ boson intermediate process, the dominant backgrounds include the $W^{+}jj$, $W^{+}b\bar{b}$, $W^{+}g \rightarrow t\bar{b}$, $bq \rightarrow tj$ and $t\bar{t}$ processes. To suppress these backgrounds, we first require a bottom quark (b-tagging) in the final jets, with tagging efficiency 0.6 and the mis-tagging efficiency neglected. Then we attempt to use various kinematics variables to highlight the excess over the standard model prediction in the observation of final states with $2~\text{jets}+1~\text{lepton}+~\met$. In the following we investigate the excluded $W'$ mass region from four strategies, i.e., the $P_T^j$-Scheme, $M_{jj}$-Scheme, $H_T$-Scheme, and $M_{t\bar{b}}$-Scheme.
\subsection{$P_T^j$-Scheme}
\begin{figure}[ht!]
    \centering
   \subfloat[ ]{
   \begin{minipage}[t]{0.4\textwidth}
   \centering
    \includegraphics[width=7cm,height=6cm]{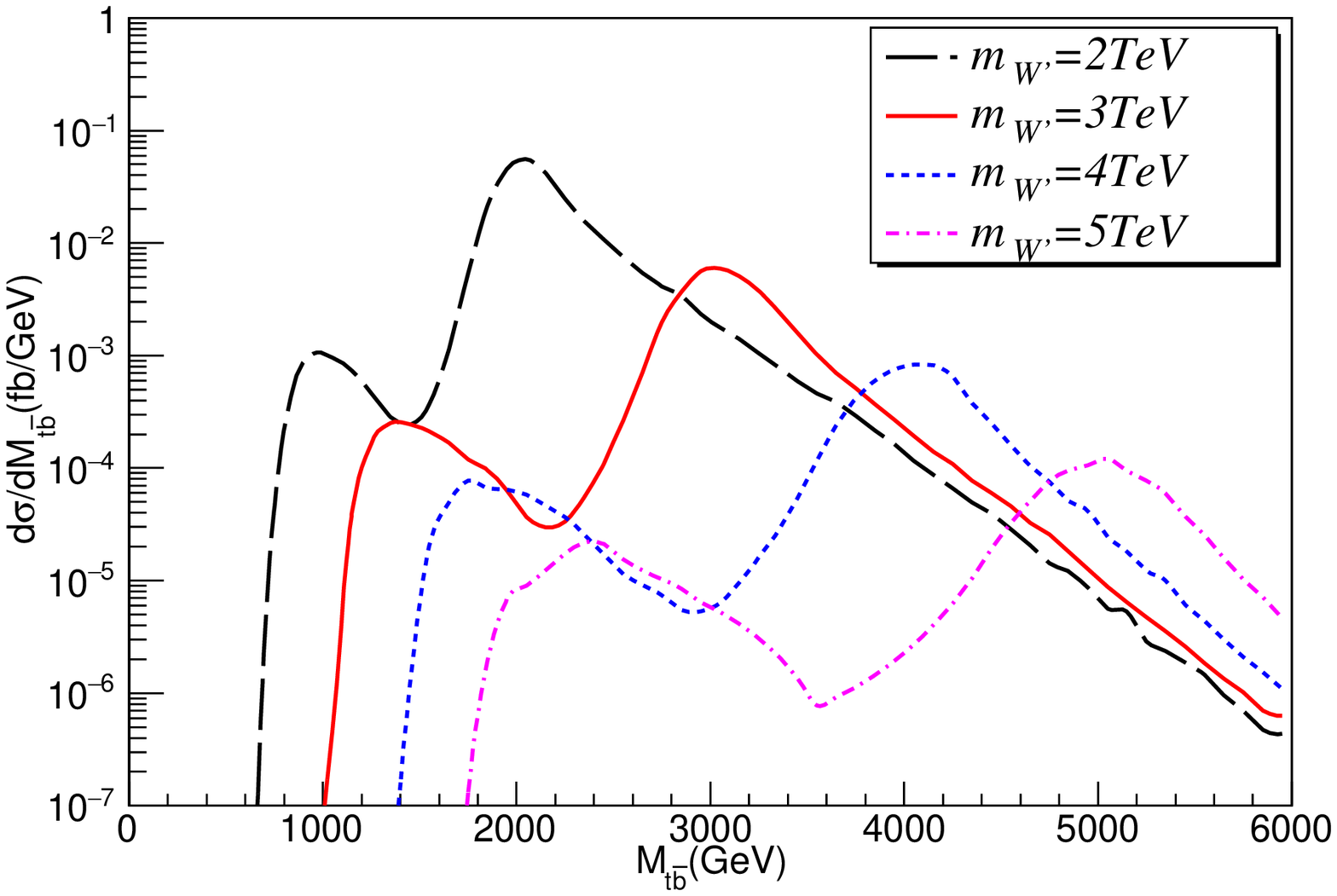}
   \end{minipage}
   }
   \subfloat[]{
   \begin{minipage}[t]{0.4\textwidth}
   \centering
   \includegraphics[width=7cm,height=6cm]{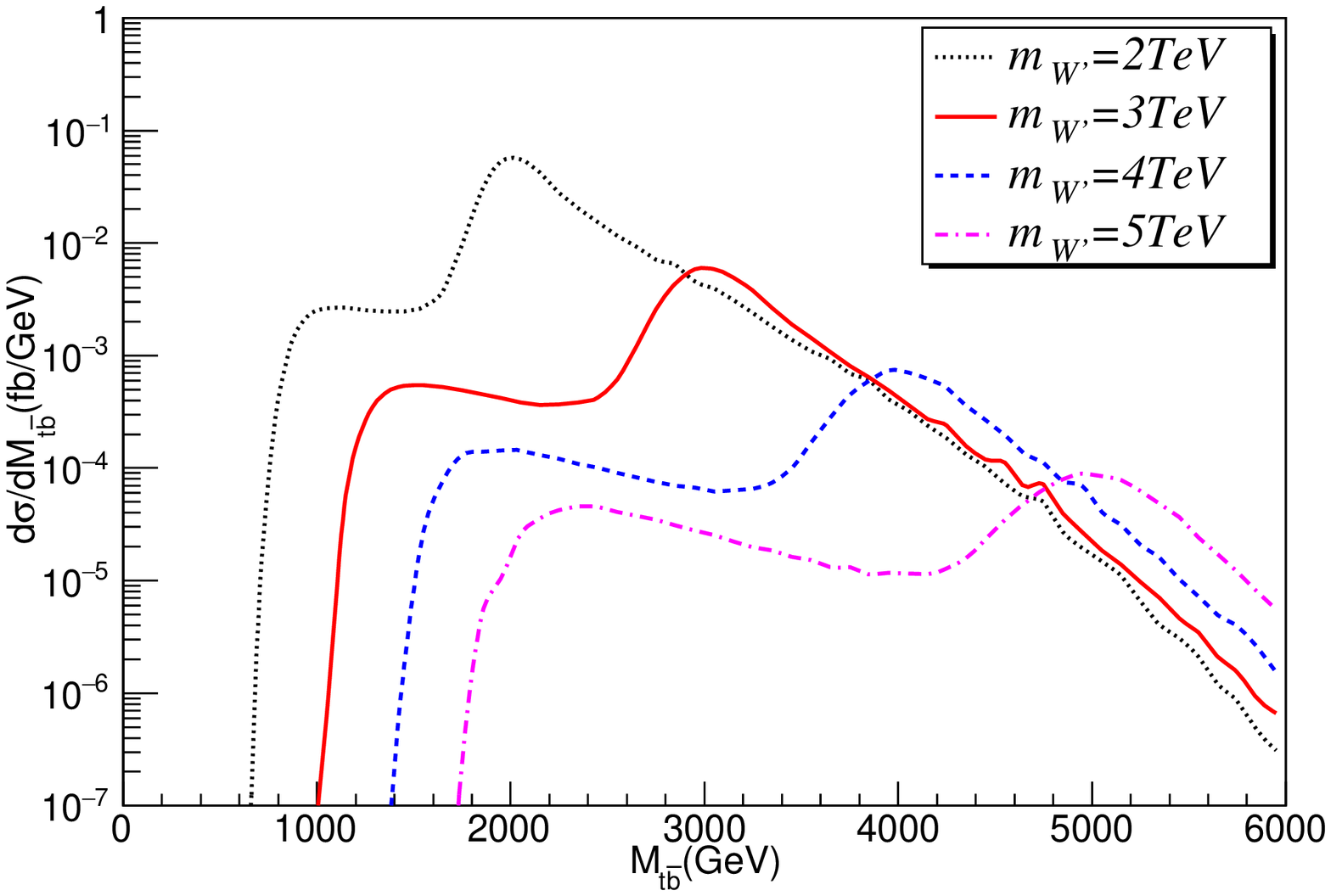}
   \end{minipage}
   }
   \caption{The invariant mass $M_{t\bar{b}}$-distribution for $m_{W^{\prime}}= 2, 3, 4, 5$ TeV at 14 TeV with the basic cuts and $P_T^{j1} > \frac{1}{5}m_{W^{\prime}}$, $P_T^{j2} > 100$ GeV for the process $ pp\rightarrow W^{\prime +} \slash  W^{+} \rightarrow \bar{b} t \rightarrow  \bar{b} b l^{+}\nu, ~~~~l^+=e^+, \mu^+$. (a) $W'=W'_{L}$; (b) $W'=W'_{R}$.} \label{fig:ptjetds}
  \end{figure}
 \begin{figure}[ht!]
    \centering
   \subfloat[]{
    \label{fig:ptjetneventL}
   \begin{minipage}[t]{0.4\textwidth}
   \centering
    \includegraphics[width=7cm,height=6cm]{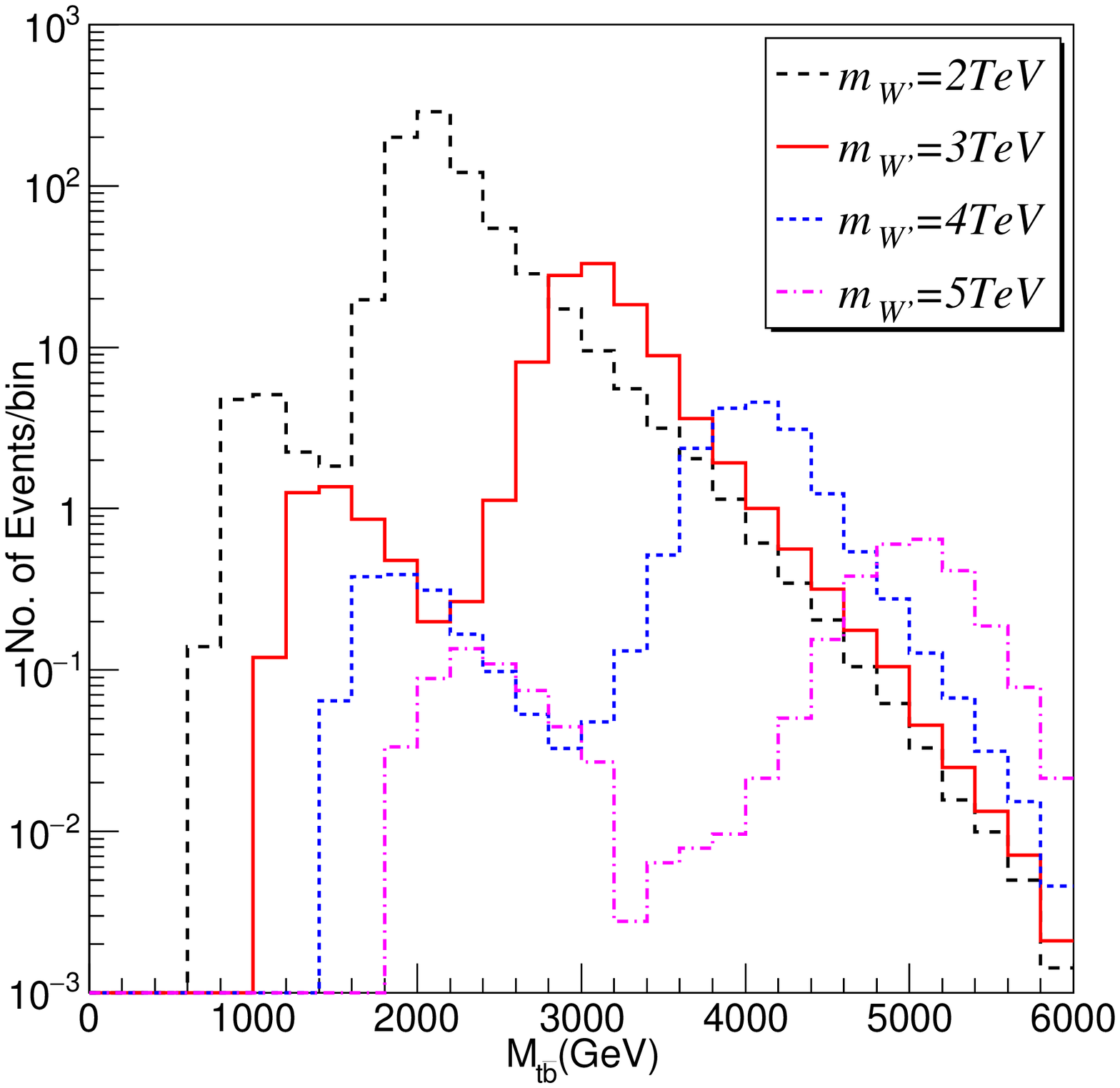}
   \end{minipage}
   }
   \subfloat[]{
   \label{fig:ptjetneventR}
   \begin{minipage}[t]{0.4\textwidth}
   \centering
   \includegraphics[width=7cm,height=6cm]{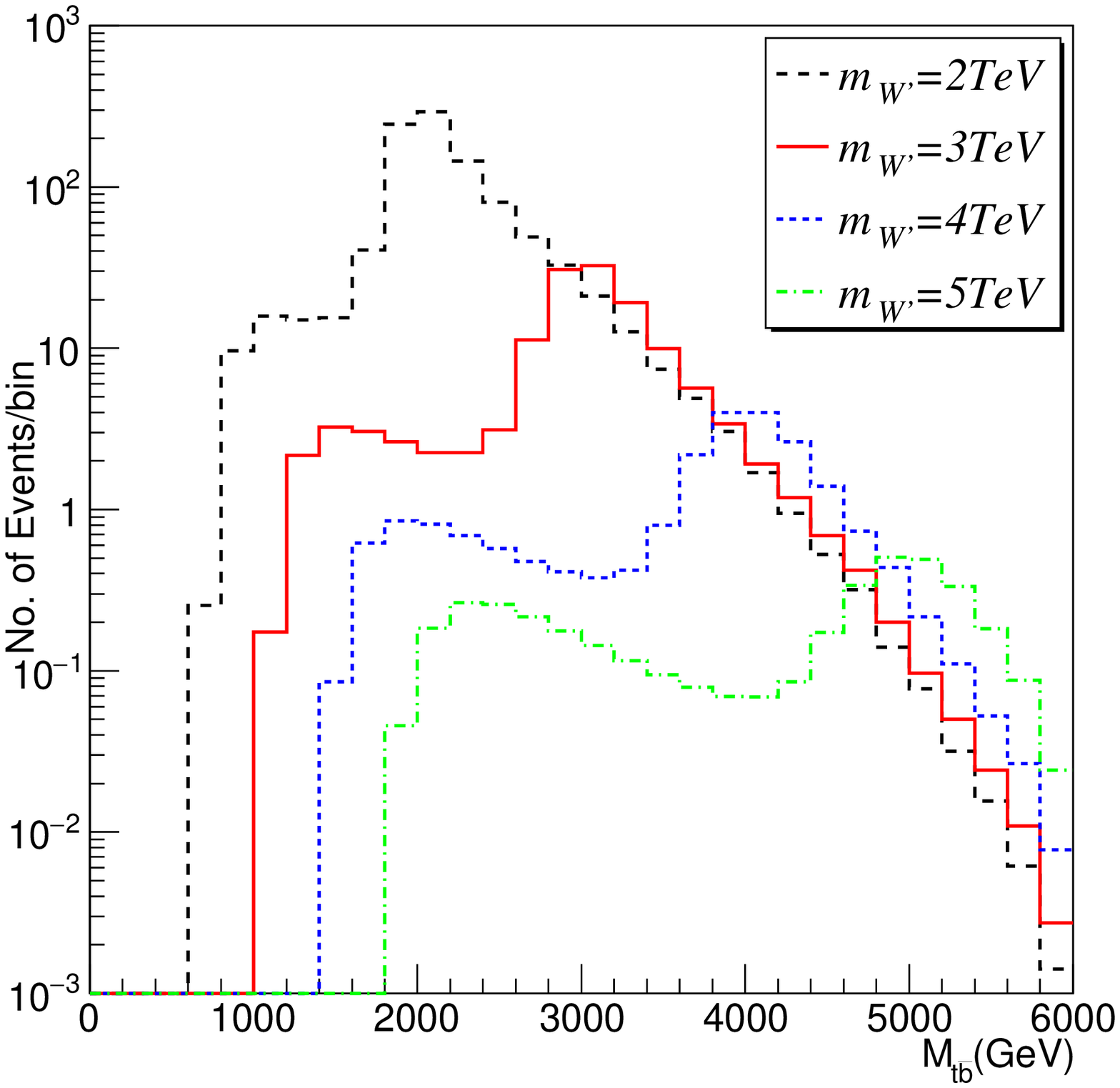}
   \end{minipage}
   }
   \caption{Number of events in each bin (200 GeV) with respect to the invariant mass $M_{t\bar{b}}$ at 14 TeV with the basic cuts and $P_T^{j1} > \frac{1}{5}m_{W^{\prime}}$, $P_T^{j2} > 100$ GeV for the process $ pp\rightarrow W^{\prime +} \slash  W^{+} \rightarrow \bar{b} t \rightarrow  \bar{b} b l^{+}\nu, ~~~~l^+=e^+, \mu^+$. (a) $W'=W'_{L}$; (b) $W'=W'_{R}$.}\label{fig:ptjetneventL}
  \end{figure}
\begin{figure}[ht!]
  \centering
  \includegraphics[width=10cm]{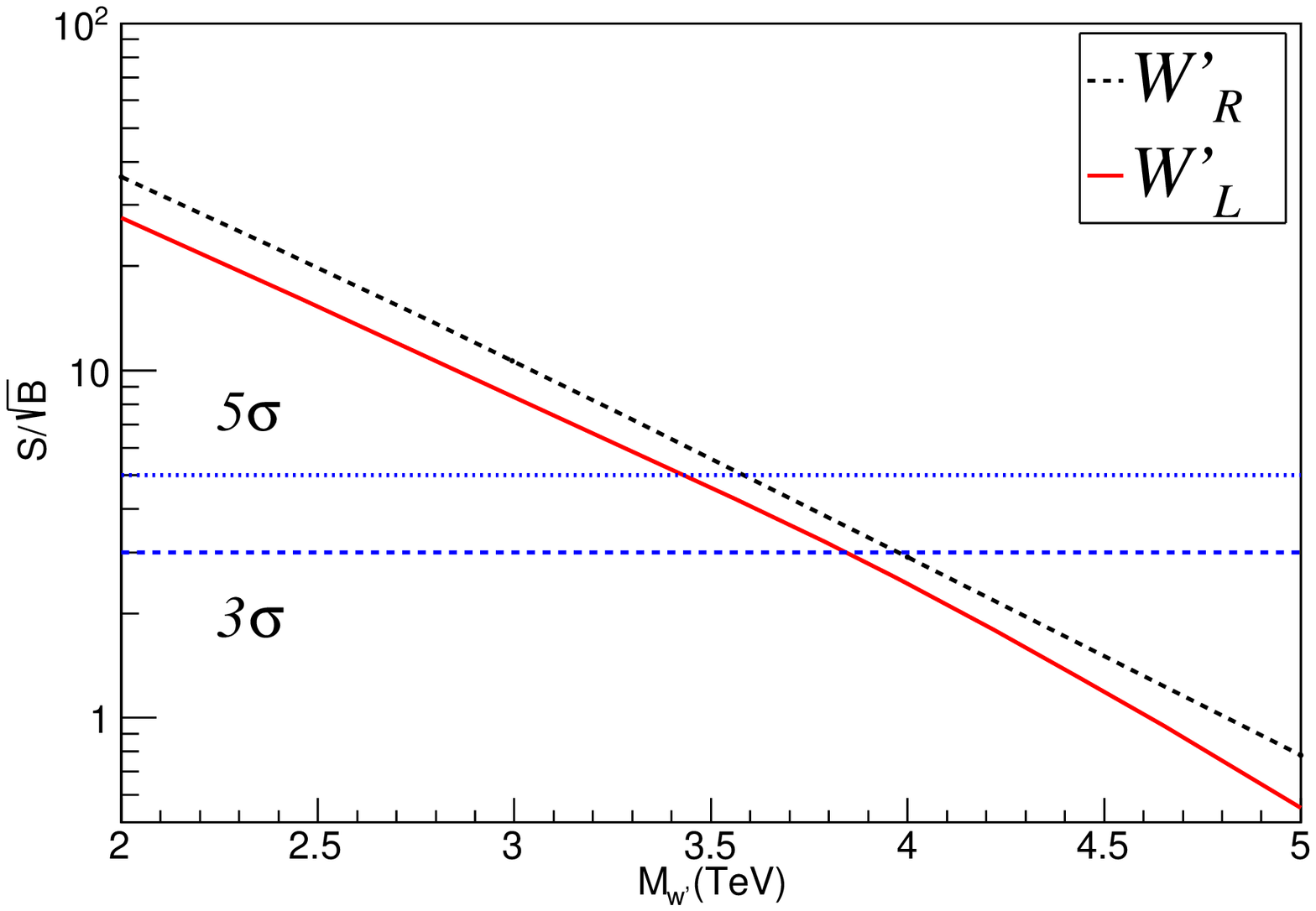}\\
  \caption{The significance distribution with different $W^{\prime}$ mass at 14 TeV with the basic cuts, $P_T^{j1} > \frac{1}{5}m_{W^{\prime}}$ and $P_T^{j2} > 100$~GeV.}\label{fig:ptjetsb}
\end{figure}

 We set cuts on the jet transverse momenta $P_T^{ji}$ (i= 1,2) with $P_T^{j1}>P_T^{j2}$,  provided the signal process has a larger number of high $P_T$ events than the $W$ boson process. Figure~\ref{fig:ptjetds} illustrates the invariant mass $M_{t\bar{b}}$-distribution for various $W'$ masses with the basic cuts and $P_T^{j1} > \frac{1}{5}m_{W^{\prime}}$, $P_T^{j2} > 100$ GeV. The lower peak in each curve is the remnant contribution from the SM $W$ boson, which is about one order of magnitude less than the signal peak. The number of events in each bin is displayed in Fig.~{\ref{fig:ptjetneventL}}. Taking $m_{W'}=3$ TeV as an example, there remain hundreds of events  after the cuts. If  we set the proper $P_T^j$ cut, the SM $W$ boson effects will be suppressed so that it would be possible to observe the excess in the $M_{t\bar{b}}$-distribution plots. The other backgrounds are investigated as well. In Table~\ref{table:background ptjet} we list the remaining cross sections after the $P_T^j$ cuts. The cross section of $W^{+}jj$ is the largest of the backgrounds, and decreases sharply with the increasing of the $P_T^{j1}$ cuts since most of the jets are soft. The background cross sections decrease with the increasing of the $P_T^{j1}$ cuts as well as the signal process, thus we adopt varying cuts of $P_T^{j1} > \frac{1}{5}m_{W^{\prime}}$ and $P_T^{j2} > 100$ GeV. The cross sections of the total background cross section and signal are listed in Table~\ref{table:significance ptjet} as well as the significance $S/\sqrt{B}$. We display the significance with respect to the $W'$ mass in Fig.~{\ref{fig:ptjetsb}} to illustrate the detectable mass region at the LHC with $\sqrt{S}=14$ TeV. It shows that the upper limit can reach 3.8 (4) TeV with a 3$\sigma$ significance for left-handed (right-handed) $W'$. Furthermore, the significance for $W'_L$ is slightly lower than $W'_R$ because of the negative effects on the cross section from the interference with the $W$ boson.
\begin{table}[h!]
\centering
\newcolumntype{d}{D{.}{.}{2}}
\begin{tabular}{|c|c|c|c|c|c|c|}
\hline
$\sigma(fb)$ & \multicolumn{1}{c|}{$W^{+}jj$(fb)} & \multicolumn{1}{c|}{$W^{+}b\bar{b}$(fb)} & \multicolumn{1}{c|}{$W^{+}g\rightarrow t\bar{b}$(fb)}& \multicolumn{1}{c|}{$bq\rightarrow tj$(fb)}& \multicolumn{1}{c|}{$t\bar{t}$} & \multicolumn{1}{c|}{$W$(fb)} \\ \hline
 $P_T^{j1}>400$ GeV &114.0&1.159&7.726&10.85&86.9&2.006\\ \hline
 $P_T^{j1}>600$ GeV  &26.44&0.2313&1.189&1.505&14.23&0.4858\\ \hline
 $P_T^{j1}>800$ GeV  &9.254&0.0608&0.1971&0.1967&1.84&0.1433 \\ \hline
 $P_T^{j1}>1000$ GeV  &3.173&0.0152&0.0435&0.0393&0.25&0.0479 \\ \hline
 $P_T^{j1}>1200$ GeV  &0&0.0005&0.0108&0.0098&0&0.0172 \\ \hline
\end{tabular}
\caption{The cross sections of SM backgrounds at 14 TeV with the basic cuts, $P_T^{j2}>100$ GeV and various $P_T^{j1}$ cuts.}\label{table:background ptjet}
\end{table}
\begin{table}[h!]
\centering
\newcolumntype{d}{D{.}{.}{2}}
\begin{tabular}{|c|c|c|c|c|c|c|c|c|c|}
\hline
 & \multicolumn{2}{c|}{$m_{W^{\prime}}=2$ TeV} & \multicolumn{2}{c|}{$m_{W^{\prime}}=3$ TeV} & \multicolumn{2}{c|}{$m_{W^{\prime}}=4$ TeV}& \multicolumn{2}{c|}{$m_{W^{\prime}}=5$ TeV} \\ \hline
  &\multicolumn{1}{c|}{$W^{\prime}_{L}$}& \multicolumn{1}{c|}{$W^{\prime}_{R}$} & \multicolumn{1}{c|}{$W^{\prime}_{L}$}& \multicolumn{1}{c|}{$W^{\prime}_{R}$} & \multicolumn{1}{c|}{$W^{\prime}_{L}$}& \multicolumn{1}{c|}{$W^{\prime}_{R}$}& \multicolumn{1}{c|}{$W^{\prime}_{L}$}& \multicolumn{1}{c|}{$W^{\prime}_{R}$} \\ \hline
 $\sigma_S(fb)$ &23.800&31.180&3.220&4.060&0.480&0.580&0.060&0.085\\ \hline
 $\sigma_B(fb)$ & \multicolumn{2}{c|}{222.64} & \multicolumn{2}{c|}{44.08} & \multicolumn{2}{c|}{11.7}& \multicolumn{2}{c|}{3.29} \\ \hline
 $S/\sqrt{B}$  &27.6&36.2&8.4&10.6&2.4&2.9&0.6&0.8\\ \hline
\end{tabular}
\caption{The cross sections of signal ($\sigma_S$) and SM backgrounds ($\sigma_B$) at 14 TeV with the basic cut, $P_T^{j1} > \frac{1}{5}m_{W^{\prime}}$ and $P_T^{j2} > 100$ GeV.} \label{table:significance ptjet}
\end{table}
\subsection{$M_{jj}$-Scheme}
\begin{figure}[ht!]
  \centering
  \includegraphics[width=10cm]{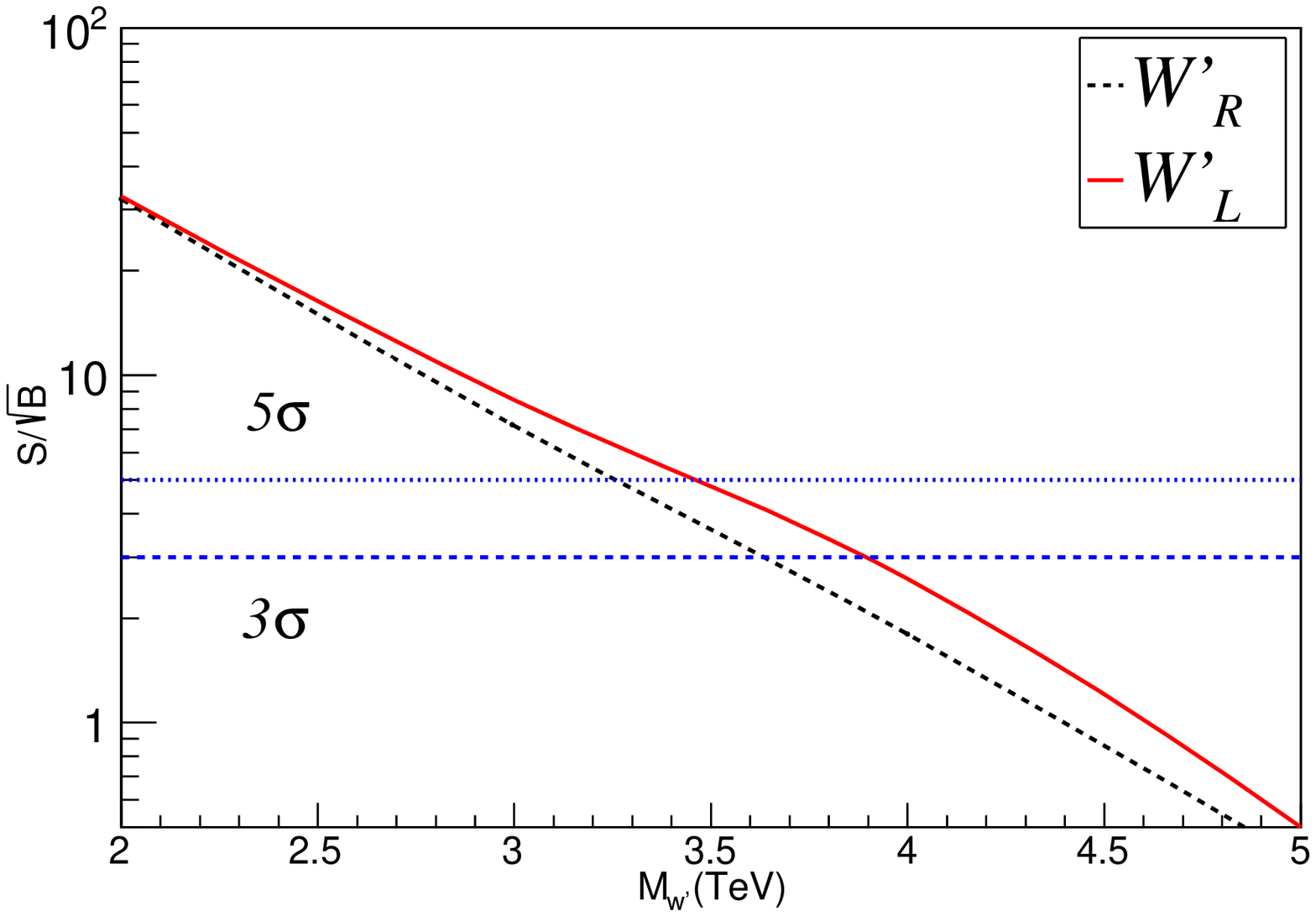}\\
  \caption{The significance distribution with different $W^{\prime}$ mass at 14 TeV with the basic cuts and $M_{jj} > \frac{1}{2} m_{W^{\prime}}$.} \label{fig:05mjjsb}
\end{figure}

 The distribution of the invariant mass of the two jets $M_{jj}$ for the signal is different from the backgrounds. We show  the number of events in each bin with respect to the invariant mass $M_{jj}$ in Fig.~{\ref{fig:mjjeventL}}, where the basic cuts are required as well as $M_{jj} > \frac{1}{2} m_{W^{\prime}}$. The influence of the $W$ boson  can be neglected in the $M_{jj}$ distribution after cuts. Compared with the $M_{t\bar{b}}$ distribution in Fig.~{\ref{fig:ptjetneventL}}, there is no clear peak in the curves, while the excess is obvious. Moreover, the  plateau is broader but lower with increasing $W^{\prime}$ mass .
\begin{figure}
    \centering
   \subfloat[]{
    \label{fig:mjjeventL}
   \begin{minipage}[t]{0.4\textwidth}
   \centering
    \includegraphics[width=7cm,height=6cm]{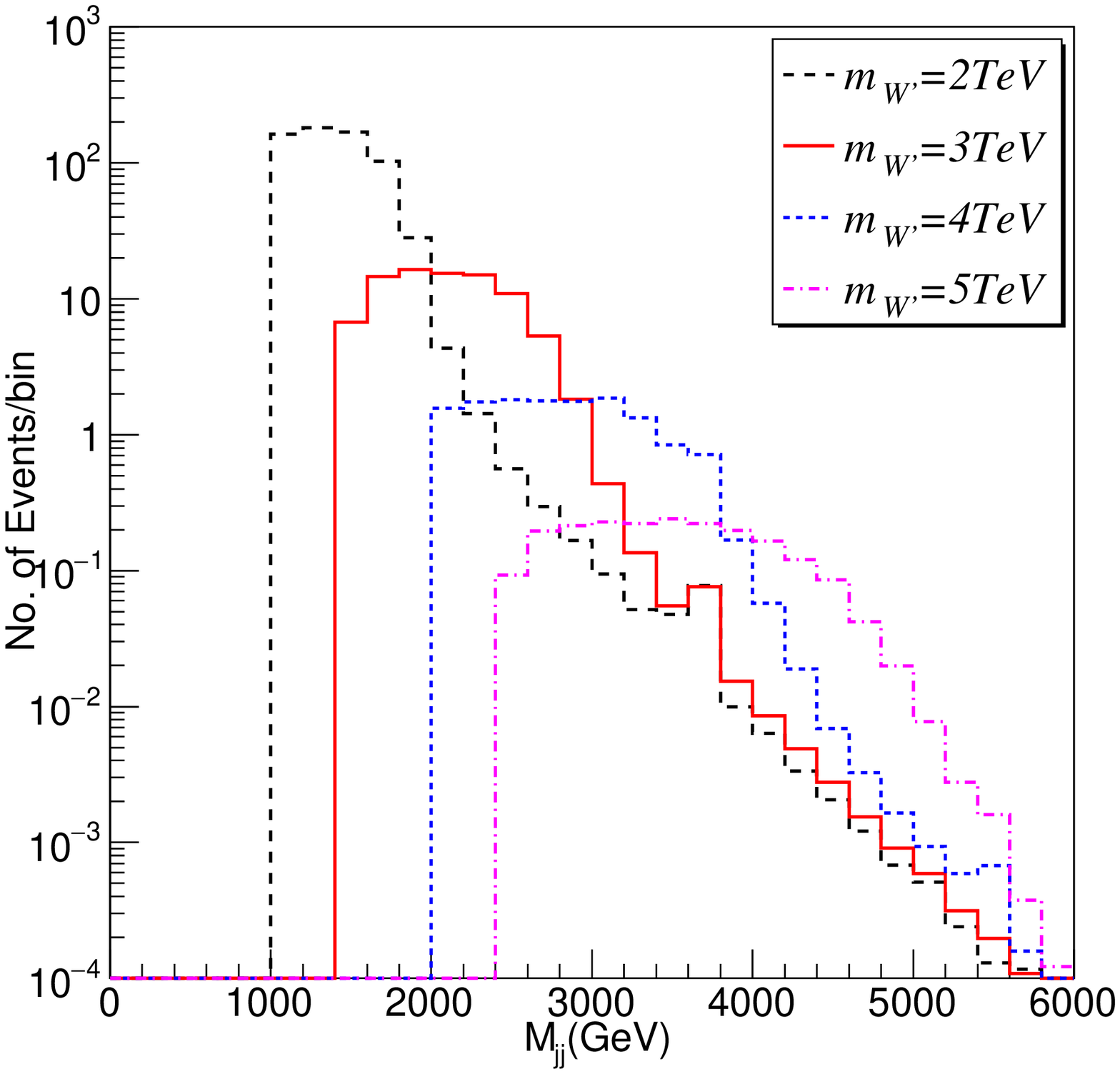}
   \end{minipage}
   }
   \subfloat[]{
   \label{fig:mjjeventR}
   \begin{minipage}[t]{0.4\textwidth}
   \centering
   \includegraphics[width=7cm,height=6cm]{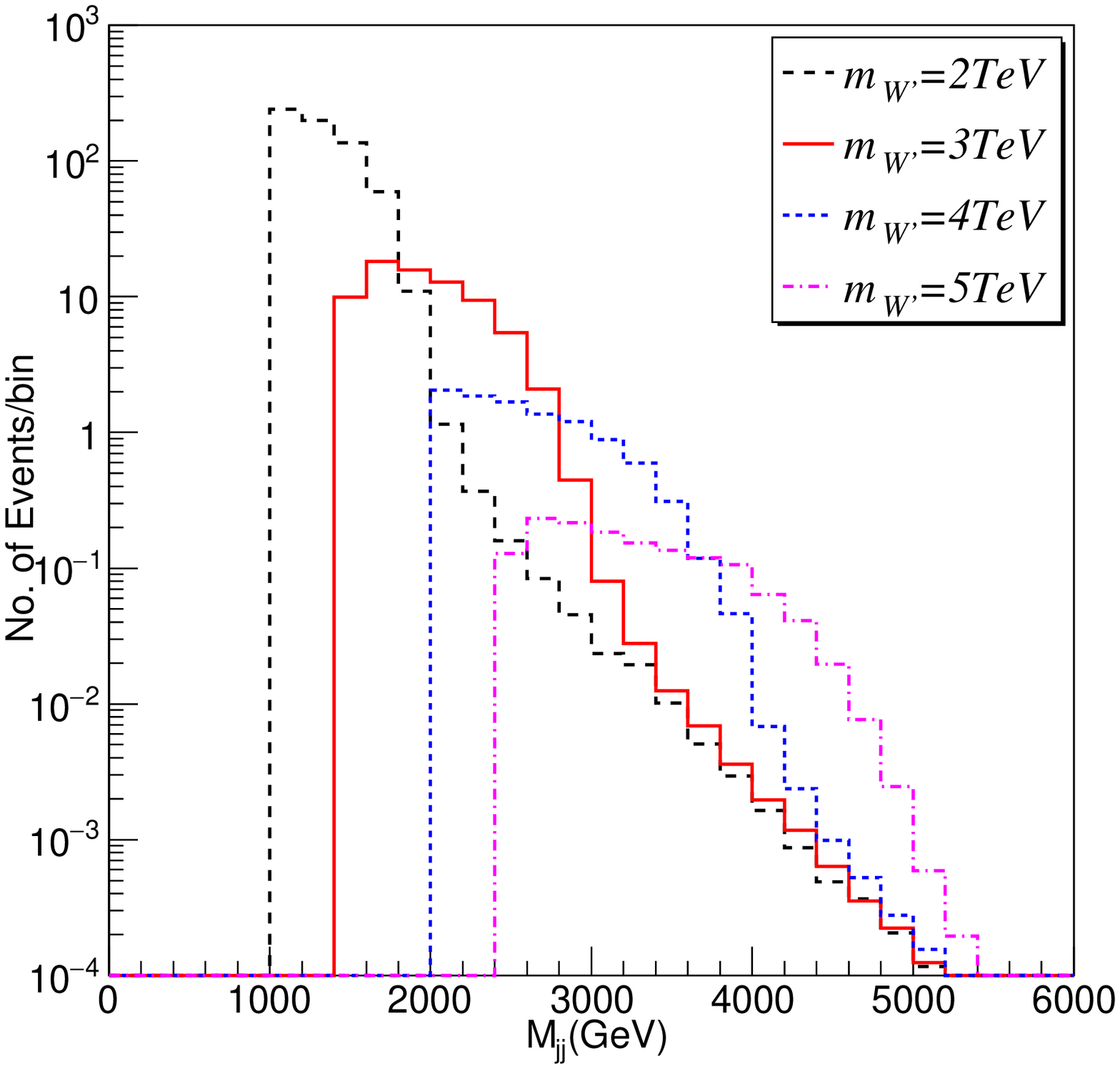}
   \end{minipage}
   }
   \caption{
   The number of events in each bin (200 GeV) with respect to the invariant mass $M_{jj}$ at 14 TeV with the basic cuts and $M_{jj}>\frac{1}{2}m_{W^{\prime}}$ for the process $ pp\rightarrow W^{\prime +} \slash  W^{+} \rightarrow \bar{b} t \rightarrow  \bar{b} b l^{+}\nu, ~~~~l^+=e^+, \mu^+$. (a) $W'=W'_{L}$, (b) $W'=W'_{R}$. }\label{fig:mjjeventL}
  \end{figure}
 The cross sections of backgrounds with the basic cuts and varying $M_{jj}$ cuts are listed in Table~{\ref{table:background mjj}} . After we set $M_{jj}>3000$ GeV, the main background is $W^+jj$, with a cross section of $0.26 fb$. The cross sections of signal and backgrounds are listed in Table~\ref{table:significance mjjcut} with different $W'$ masses. Supposing the $W'$ mass is 4 TeV, after we set a cut of $M_{jj}>2000$ GeV, there remain 132 (93) events for $W'_L$ ($W'_R$) at 14 TeV LHC with luminosity of $300 fb^{-1}$.
\begin{table}[h!]
\centering
\newcolumntype{d}{D{.}{.}{2}}
\begin{tabular}{|c|c|c|c|c|c|c|}
\hline
$\sigma(fb)$ & \multicolumn{1}{c|}{$W^{+}jj$} & \multicolumn{1}{c|}{$W^{+}b\bar{b}$} & \multicolumn{1}{c|}{$W^{+}g\rightarrow t\bar{b}$}& \multicolumn{1}{c|}{$bq\rightarrow tj$}& \multicolumn{1}{c|}{$t\bar{t}$} & \multicolumn{1}{c|}{$W$} \\ \hline
 $M_{jj}>1000$~GeV &104.7&0.4459&13.19&1.495&8.710&0.506\\ \hline
 $M_{jj}>1500$~GeV &28.56&0.0725&2.952&0.2065&1.090&0.084\\ \hline
 $M_{jj}>2000$~GeV &7.932&0.0018&0.7667&0.0197&0.080&0.0184\\ \hline
 $M_{jj}>2500$~GeV &3.173&0.0039&0.2281&0.0098&0&0.0468 \\ \hline
 $M_{jj}>3000$~GeV &0.2644&0.0010&0.0590&0&0&0.0013 \\ \hline
\end{tabular}
\caption{The cross sections of SM backgrounds at 14 TeV with basic cuts and $M_{jj}$ cut.}\label{table:background mjj}
\end{table}
\begin{table}[h!]
\centering
\newcolumntype{d}{D{.}{.}{2}}
\begin{tabular}{|c|c|c|c|c|c|c|c|c|c|}
\hline
 & \multicolumn{2}{c|}{$m_{W^{\prime}}=2$ TeV} & \multicolumn{2}{c|}{$m_{W^{\prime}}=3$ TeV} & \multicolumn{2}{c|}{$m_{W^{\prime}}=4$ TeV}& \multicolumn{2}{c|}{$m_{W^{\prime}}=5$ TeV} \\ \hline
 & \multicolumn{1}{c|}{$W^{\prime}_{L}$}& \multicolumn{1}{c|}{$W^{\prime}_{R}$} & \multicolumn{1}{c|}{$W^{\prime}_{L}$}& \multicolumn{1}{c|}{$W^{\prime}_{R}$} & \multicolumn{1}{c|}{$W^{\prime}_{L}$}& \multicolumn{1}{c|}{$W^{\prime}_{R}$}& \multicolumn{1}{c|}{$W^{\prime}_{L}$}& \multicolumn{1}{c|}{$W^{\prime}_{R}$} \\ \hline
 $\sigma_S(fb)$ &21.420&21.100&2.831&2.400&0.440&0.310&0.050&0.042\\ \hline
 $\sigma_B(fb)$ & \multicolumn{2}{c|}{129.04} & \multicolumn{2}{c|}{32.97} & \multicolumn{2}{c|}{8.81}& \multicolumn{2}{c|}{3.4} \\ \hline
 $S/\sqrt{B}$  &32.7&33.2&8.5&7.2&2.6&1.8&0.5&0.4\\ \hline
\end{tabular}
\caption{The cross sections of signal ($\sigma_S$) and SM backgrounds ($\sigma_B$) at 14 TeV with the basic cuts and $M_{jj}>\frac{1}{2}m_{W^{\prime}}$.}\label{table:significance mjjcut}
\end{table}
Figure~\ref{fig:05mjjsb} illustrates the detectable $W'$ mass region at 14 TeV with the basic cuts and $M_{jj} > \frac{1}{2} m_{W^{\prime}}$ for $S/\sqrt{B}>3$. The $W'$ mass should be larger than 3.6 (3.9) TeV with a 3$\sigma$ significance for $W'_R$ ($W'_L$) if there is no excess in the $M_{jj}$ distribution.
\subsection{$H_T$-Scheme}
\begin{figure}[h!]
    \centering
   \subfloat[]{
   \begin{minipage}[t]{0.4\textwidth}
   \centering
    \includegraphics[width=7cm,height=6cm]{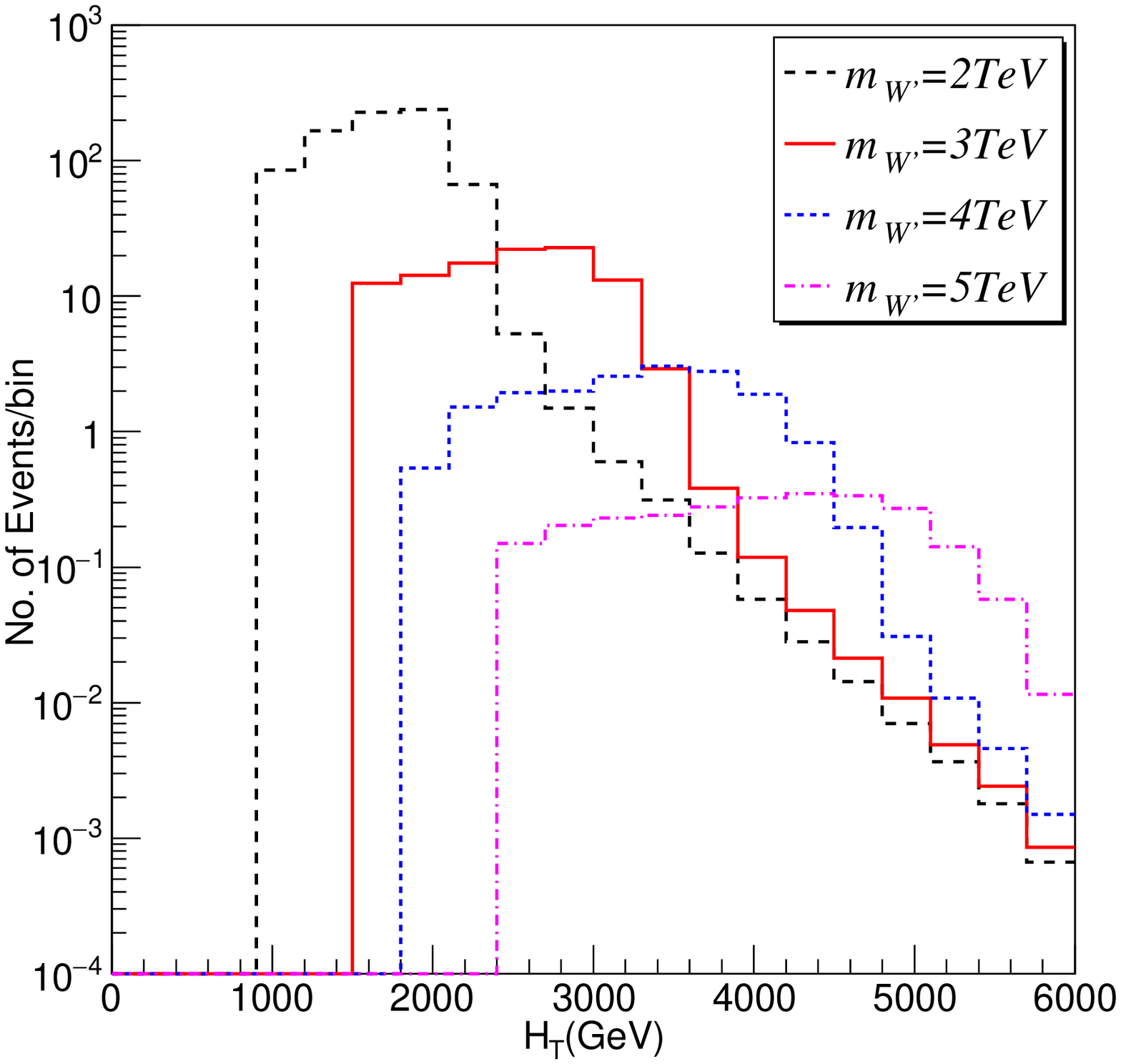}
   \end{minipage}
   }
   \subfloat[]{
   \begin{minipage}[t]{0.4\textwidth}
   \centering
   \includegraphics[width=7cm,height=6cm]{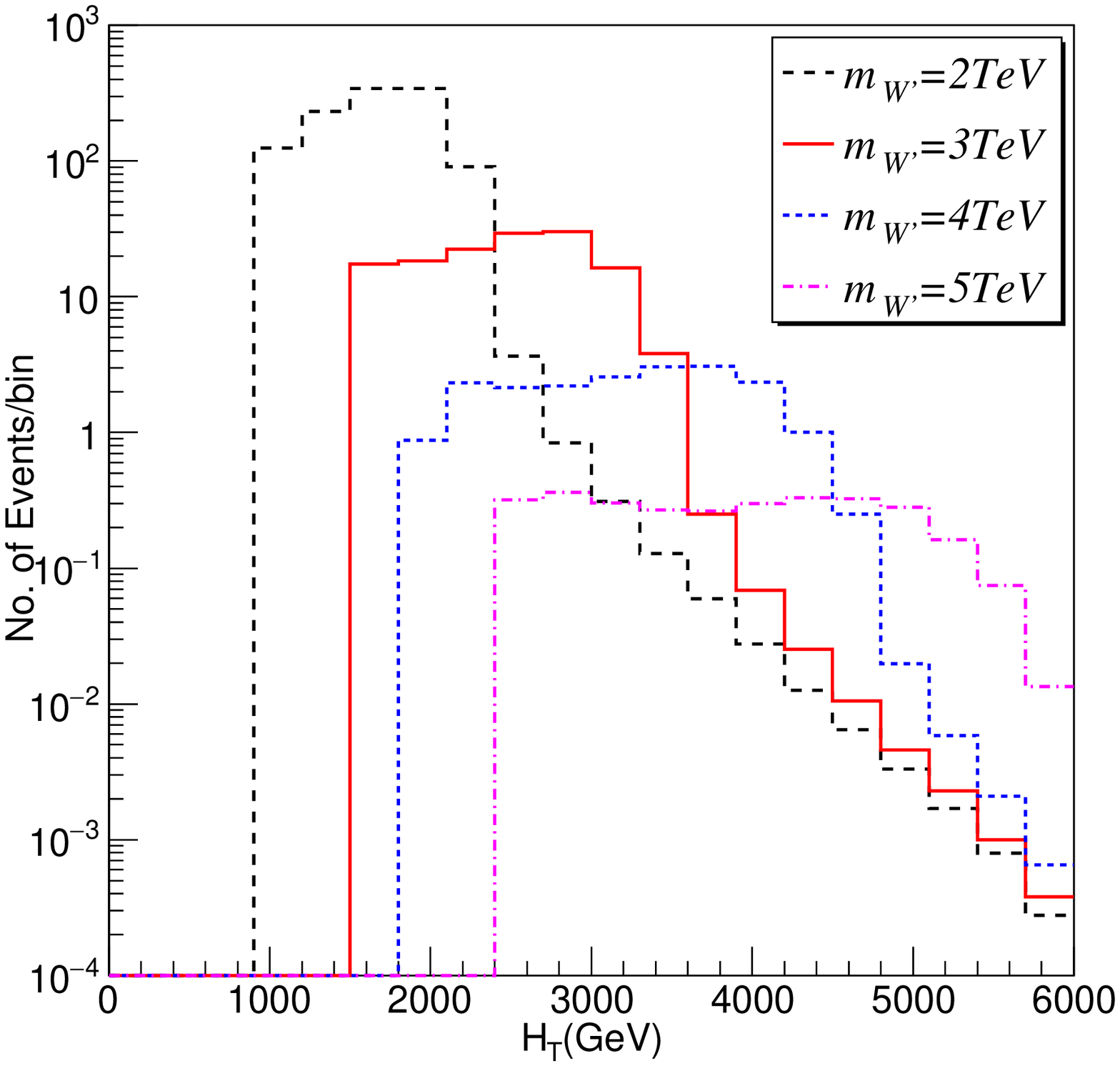}
   \end{minipage}
   }
   \caption{
   The number of events in each bin (200 GeV) with respect to the collision energy scale $H_T$ at 14 TeV with the basic cuts and $H_T>\frac{1}{2}m_{W^{\prime}}$ for the process $ pp\rightarrow W^{\prime +} \slash  W^{+} \rightarrow \bar{b} t \rightarrow  \bar{b} b l^{+}\nu, ~~~~l^+=e^+, \mu^+$. (a) $W'=W'_{L}$; (b) $W'=W'_{R}$. }\label{fig:htneventds}
\end{figure}

\begin{figure}[ht!]
  \centering
  \includegraphics[width=10cm]{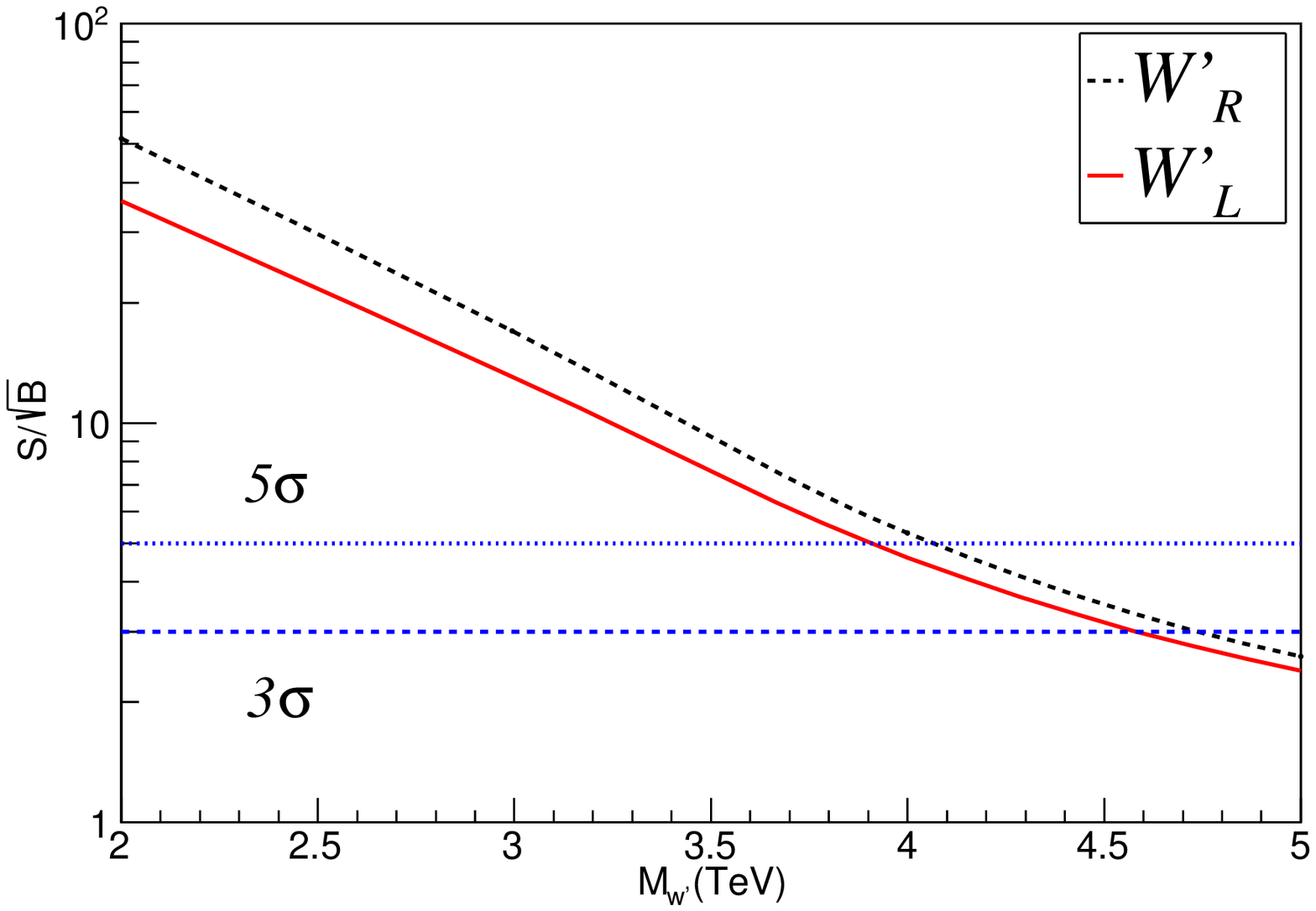}\\
  \caption{The significance distribution for different $W^{\prime}$ mass at 14 TeV with the basic cuts and $H_T > \frac{1}{2} m_{W^{\prime}} $. }\label{fig:htsiba}
\end{figure}

 Due to the large mass of the $W'$ boson, the signal process can happen only if a lot of energy is transferred in the collision. Thus we can use a high energy scale $H_T$ to distinguish the signal and backgrounds.  $H_T$ is the scalar sum of the transverse momentum for the final state, which is defined as
\begin{align}
H_T=P_T^{j1}+P_T^{j2}+P_T^{\ell}+\met,
\end{align}
 Figure~\ref{fig:htneventds} shows the number of events per bin with respect to $H_T$ with the basic cuts and $H_t>\frac{1}{2}m_{W'}$.  It has a broad plateau in each curve, like in the $M_{jj}$ distribution, while the upper mass limit for $W'$ is up to 5 TeV for twenty events remaining.
The cross sections of backgrounds  are listed in Table~{\ref{table:background Ht}} with the basic cuts and varying $M_{jj}$ cut.  As shown in the table, the $W^+jj$ and $bq\to tj$ processes are cut down to zero after the $H_T> 3000$ GeV cut, while other backgrounds have a tiny cross section left. A suitable$H_T$ cut is therefore an effective way to suppress the SM backgrounds.
\begin{table}[h!]
\centering
\newcolumntype{d}{D{.}{.}{2}}
\begin{tabular}{|c|c|c|c|c|c|c|}
\hline
$\sigma(fb)$ & \multicolumn{1}{c|}{$W^{+}jj$} & \multicolumn{1}{c|}{$W^{+}b\bar{b}$} & \multicolumn{1}{c|}{$W^{+}g\rightarrow t\bar{b}$}& \multicolumn{1}{c|}{$bq\rightarrow tj$}& \multicolumn{1}{c|}{$t\bar{t}$} & \multicolumn{1}{c|}{$W$} \\ \hline
 $H_T>1000$~GeV  &85.4&0.9222&3.734&5.035&54&1.276\\ \hline
 $H_T>1500$~GeV  &14.81&0.1490&0.3508&0.354&4.355&0.240\\ \hline
 $H_T>2000$~GeV  &3.437&0.0294&0.0357&0.0393&0.17&0.060 \\ \hline
 $H_T>2500$~GeV  &0.2644&0.0074&0.0109&0.0197&0&0.017 \\ \hline
 $H_T>3000$~GeV  &0&0.0020&0.0031&0&0&0.005 \\ \hline
 $H_T>3500$~GeV  &0&0.0010&0.0031&0&0&0.0012 \\ \hline
\end{tabular}
\caption{The cross sections of SM backgrounds at 14 TeV with the basic cuts and $H_T$ cut.}\label{table:background Ht}
\end{table}
\begin{table}[h!]
\centering
\newcolumntype{d}{D{.}{.}{2}}
\begin{tabular}{|c|c|c|c|c|c|c|c|c|c|}
\hline
 & \multicolumn{2}{c|}{$m_{W^{\prime}}=2$ TeV} & \multicolumn{2}{c|}{$m_{W^{\prime}}=3$ TeV} & \multicolumn{2}{c|}{$m_{W^{\prime}}=4$ TeV}& \multicolumn{2}{c|}{$m_{W^{\prime}}=5$ TeV} \\ \hline
 & \multicolumn{1}{c|}{$W^{\prime}_{L}$}& \multicolumn{1}{c|}{$W^{\prime}_{R}$} & \multicolumn{1}{c|}{$W^{\prime}_{L}$}& \multicolumn{1}{c|}{$W^{\prime}_{R}$} & \multicolumn{1}{c|}{$W^{\prime}_{L}$}& \multicolumn{1}{c|}{$W^{\prime}_{R}$}& \multicolumn{1}{c|}{$W^{\prime}_{L}$}& \multicolumn{1}{c|}{$W^{\prime}_{R}$} \\ \hline
 $\sigma_S(fb)$ &25.490&36.550&3.380&4.360&0.520&0.590&0.076&0.083\\ \hline
 $\sigma_B(fb)$ & \multicolumn{2}{c|}{150.36} & \multicolumn{2}{c|}{20.26} & \multicolumn{2}{c|}{3.77}& \multicolumn{2}{c|}{0.31} \\ \hline
 $S/\sqrt{B}$  &36.0&51.6&13.0&16.8&4.6&5.3&2.4&2.6\\ \hline
\end{tabular}
\caption{The cross sections of signal ($\sigma_S$) and SM backgrounds ($\sigma_B$) and the significance at 14 TeV with the basic cuts and $H_T> \frac{1}{2}M_{W^{\prime}}$.}\label{table:significance HTcut}
\end{table}
 The total cross sections for signal and backgrounds are summarized in Table~\ref{table:significance HTcut}. For $m_{W'}=4$ TeV, there are about 156 (177) events for the $W'_L$ ($W'_R$) process and 1080 events for backgrounds with a cut of $H_T > \frac{1}{2} m_{W^{\prime}}$. As shown in Fig.~\ref{fig:htsiba}, the $W'$ can be detectable with mass below 4.5 TeV in the $H_T$ distribution for $3\sigma$ significance at 14 TeV.
\subsection{ $M_{t\bar{b}}$-Scheme}
\begin{figure}
    \centering
   \subfloat[]{
    \label{fig:72fb1314sig}
   \begin{minipage}[t]{0.4\textwidth}
   \centering
    \includegraphics[width=7cm,height=6cm]{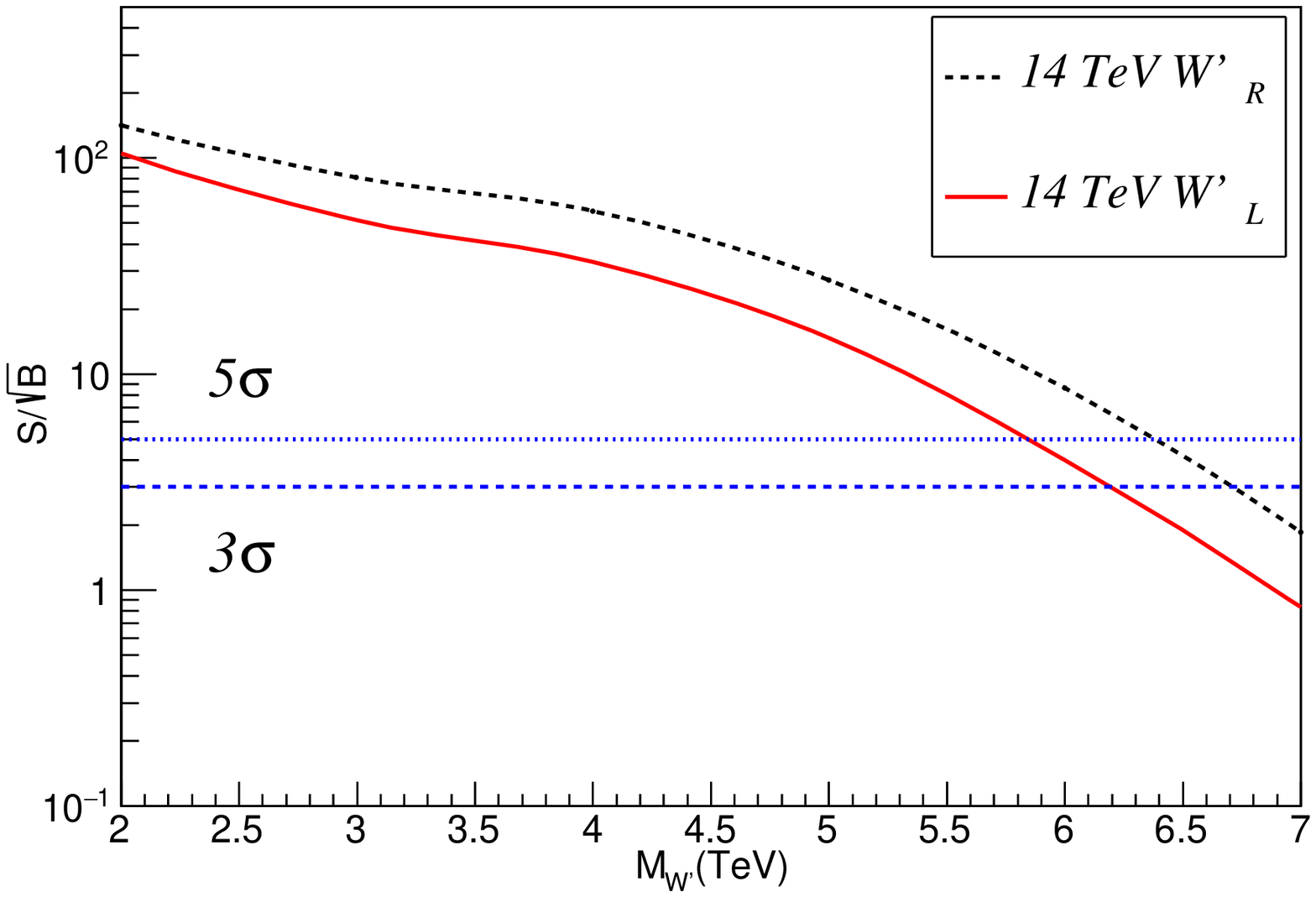}
   \end{minipage}
   }
   \subfloat[]{
   \label{fig:72fb13tevsig}
   \begin{minipage}[t]{0.4\textwidth}
   \centering
   \includegraphics[width=7cm,height=6cm]{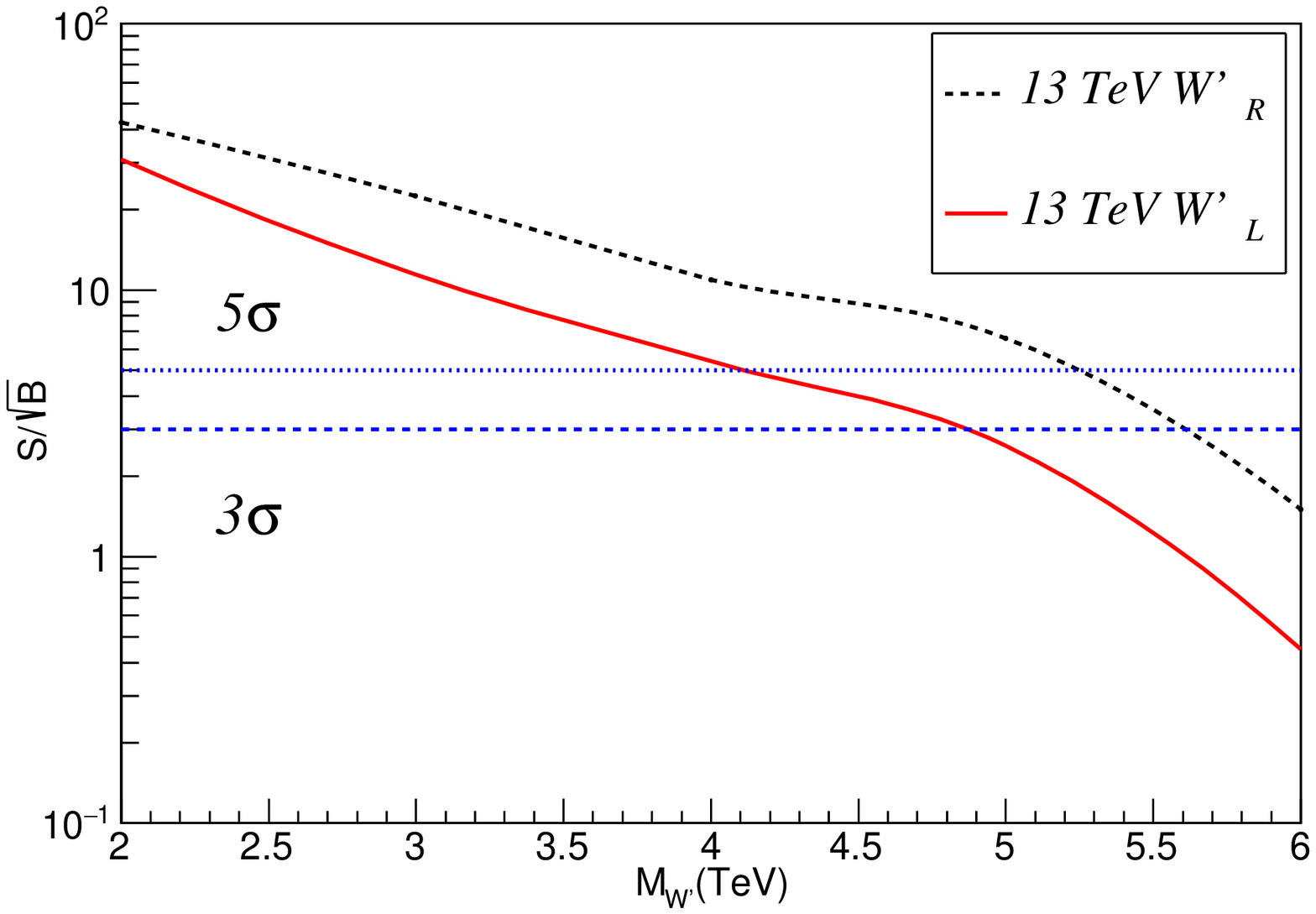}
   \end{minipage}
   }
   \caption{The significance distribution for different $W^{\prime}$ mass at the LHC with the basic cuts and $M_{t \bar b}$ cut: (a) with $M_{t\bar{b}}> \frac{3}{4}m_{W^{\prime}}$ at 14 TeV for a luminosity of $300~fb^{-1}$; (b) with $M_{t\bar{b}}> \frac{2}{3}m_{W^{\prime}}$ at 13 TeV for a luminosity of $72~fb^{-1}$.
    }
  \end{figure}

 In the $W'\to t \bar{b}$ channel, the most effective way to reconstruct the $W'$ mass peak is by the momentum of the top and bottom quarks. Provided the  top quarks decay semi-leptonically, all the momenta of the final state can be detected in the detector, except the neutrino. However, we can obtain the transverse momentum of neutrino from the conservation of transverse momentum, using the formula
  \begin{align}
  \textbf{P}_{\nu T}=-\big(\textbf{P}_{\ell T}+\textbf{P}_{j1}+\textbf{P}_{j2}\big),
  \end{align}
 where $\textbf{P}_{j T}$ is the transverse momentum of particle $j$. While its longitudinal momentum cannot be detected, we can obtain it by solving the  equation
  \begin{align}
  m_{W}^2=(P_{\nu}+P_{\ell})^2,
  \end{align}
 which implies the neutrino and charged lepton are generated by an on-shell $W$ boson. Solving this quadratic equation for the neutrino longitudinal momentum leads to a twofold ambiguity. Furthermore, we can use the solution to reconstruct the top quark invariant mass through
   \begin{align}
  M_{rt}&=\sqrt{\big(P_{\nu}+P_{\ell}+P_j\big)^2}.
  \end{align}
 We adopt cuts on the top reconstruction of
   \begin{align}
   |M_{rt}-m_t|\leq 20 ~\text{GeV}.
   \end{align}
 Provided that all the final state momenta are confirmed, then we can reconstruct the whole process. The invariant mass of $M_{tb}$ can be obtained from
   \begin{align}
   M_{t\bar{b}}&=\sqrt{\big(P_{\nu}+P_{\ell}+P_{j1}+P_{j2}\big)^2}.
   \end{align}
\begin{figure}
    \centering
   \subfloat[]{
   \begin{minipage}[t]{0.4\textwidth}
   \centering
    \includegraphics[width=7cm,height=6cm]{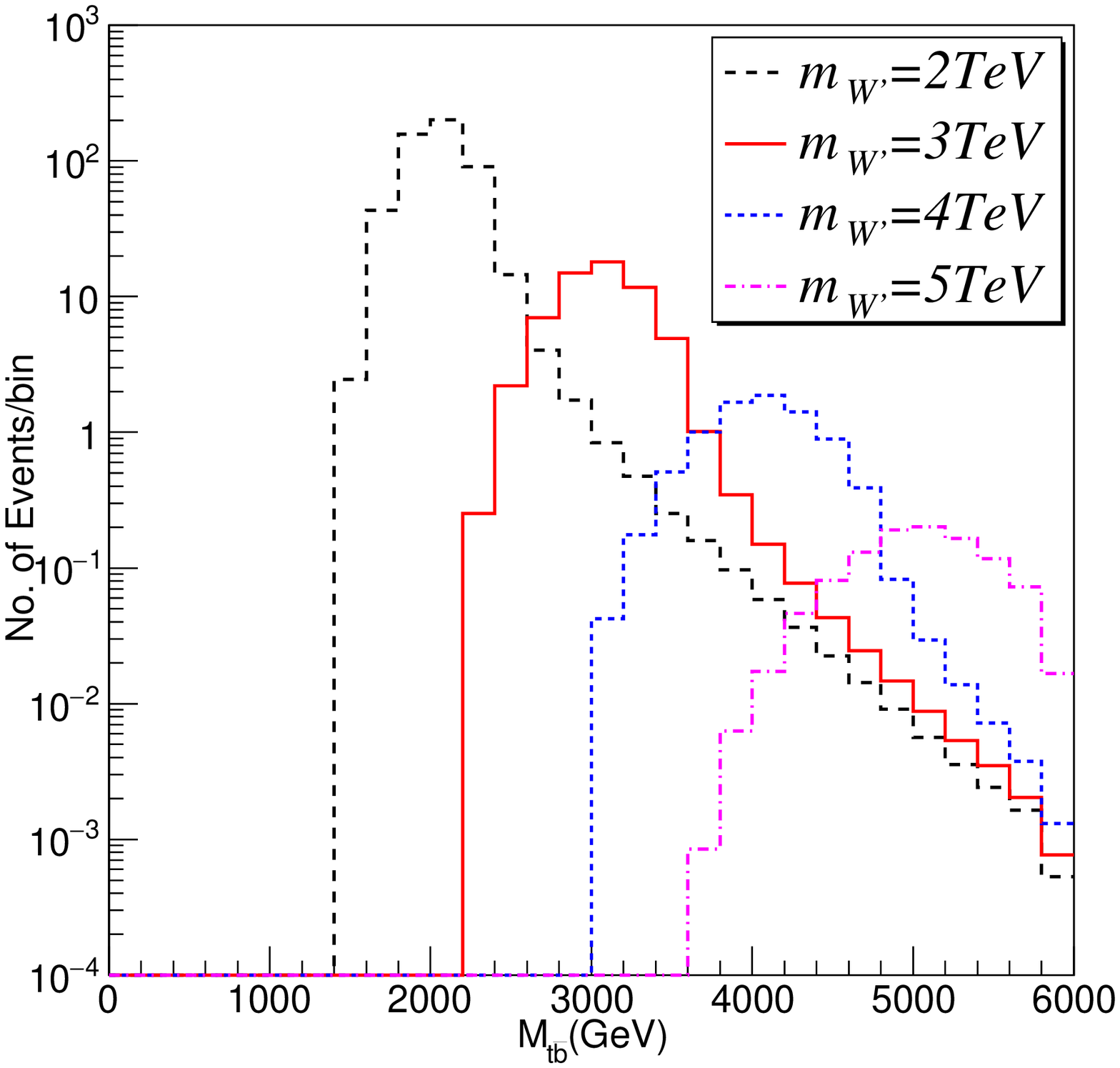}
   \end{minipage}
   }
   \subfloat[]{
   \begin{minipage}[t]{0.4\textwidth}
   \centering
   \includegraphics[width=7cm,height=6cm]{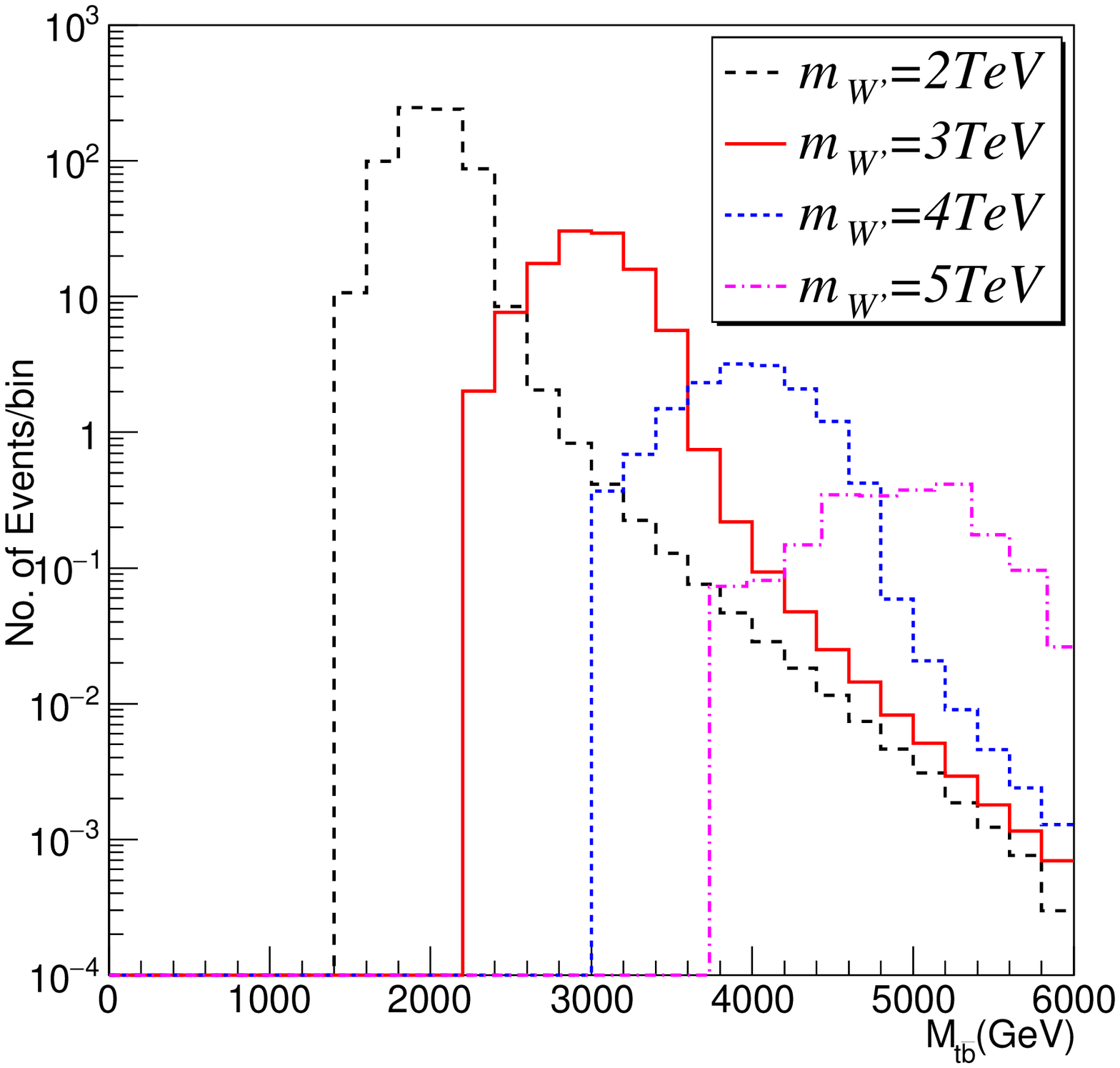}
   \end{minipage}
   }
   \caption{The number of events in each bin (200 GeV) with respect to the invariant mass $M_{t\bar{b}}$  at 14 TeV with the basic cuts and $M_{t\bar{b}}>\frac{3}{4}m_{W^{\prime}}$ for the process $ pp\rightarrow W^{\prime +} \slash  W^{+} \rightarrow \bar{b} t \rightarrow  \bar{b} b l^{+}\nu, ~~~~l^+=e^+, \mu^+$. (a) $W'=W'_{L}$; (b) $W'=W'_{R}$.}\label{fig:recontevent}
  \end{figure}
 Figure~\ref{fig:recontevent} displays the number of events per bin with basic cuts and $M_{t\bar{b}}>\frac{3}{4}M_{W'}$ for $W^{\prime}$ mass varying from 2 to 5 TeV. It is easy to find that the mass peak is clear in the $M_{t\bar{b}}$ distribution due to the whole process reconstruction. 
\begin{table}[h!]
\centering
\newcolumntype{d}{D{.}{.}{2}}
\begin{tabular}{|c|c|c|c|c|c|c|}
\hline
$\sigma(fb)$ & \multicolumn{1}{c|}{$W^{+}jj$} & \multicolumn{1}{c|}{$W^{+}b\bar{b}$} & \multicolumn{1}{c|}{$W^{+}g\rightarrow t\bar{b}$}& \multicolumn{1}{c|}{$bq\rightarrow tj$}& \multicolumn{1}{c|}{$t\bar{t}$} & \multicolumn{1}{c|}{$W$}\\ \hline
 $m_{W^{\prime}}=2$ TeV &4.495&0.0034&0.4454&0.9146 &0.3498&1.675\\ \hline
 $m_{W^{\prime}}=3$ TeV &0.2644&0&0.0372&0.0197 &0.0486&0.083\\ \hline
 $m_{W^{\prime}}=4$ TeV &0&0&0.0016&0.0098&0.0085&0 \\ \hline
 $m_{W^{\prime}}=5$ TeV &0&0&0&0&0.0016&0 \\ \hline
 $m_{W^{\prime}}=6$ TeV &0&0&0&0&0.0003&0 \\ \hline
\end{tabular}
\caption{The cross sections of SM backgrounds  at 14 TeV with reconstruction and $M_{t\bar{b}}>\frac{3}{4}m_{W'}$.}\label{table:background recont}
\end{table}
\begin{table}[h!]
\centering
\newcolumntype{d}{D{.}{.}{2}}
\begin{tabular}{|c|c|c|c|c|c|c|c|c|c|c|c|}
\hline
 & \multicolumn{2}{c|}{$m_{W^{\prime}}=2$ TeV} & \multicolumn{2}{c|}{$m_{W^{\prime}}=3$ TeV} & \multicolumn{2}{c|}{$m_{W^{\prime}}=4$ TeV}& \multicolumn{2}{c|}{$m_{W^{\prime}}=5$ TeV}& \multicolumn{2}{c|}{$m_{W^{\prime}}=6$ TeV} \\ \hline
 & \multicolumn{1}{c|}{$W^{\prime}_{L}$}& \multicolumn{1}{c|}{$W^{\prime}_{R}$} & \multicolumn{1}{c|}{$W^{\prime}_{L}$}& \multicolumn{1}{c|}{$W^{\prime}_{R}$} & \multicolumn{1}{c|}{$W^{\prime}_{L}$}& \multicolumn{1}{c|}{$W^{\prime}_{R}$}& \multicolumn{1}{c|}{$W^{\prime}_{L}$}& \multicolumn{1}{c|}{$W^{\prime}_{R}$}& \multicolumn{1}{c|}{$W^{\prime}_{L}$}&\multicolumn{1}{c|}{$W^{\prime}_{R}$} \\ \hline
 $\sigma_S(fb)$ &17.002&22.910&2.000&3.127&0.268&0.452&0.034&0.063&0.004&0.0086 \\ \hline
 $\sigma_B(fb)$ & \multicolumn{2}{c|}{7.87} & \multicolumn{2}{c|}{0.453} & \multicolumn{2}{c|}{0.02}& \multicolumn{2}{c|}{0.0016}&\multicolumn{2}{c|}{0.0003} \\ \hline
 $S/\sqrt{B}$  &105.0&141.4&51.5&80.5&33.0&56.8&14.7&27.3&4.0&8.6 \\ \hline
\end{tabular}
\caption{The cross sections of signal ($\sigma_S$) and SM backgrounds ($\sigma_B$) at 14 TeV with basic cut, reconstruction and $M_{t\bar{b}}>\frac{3}{4}m_{W'}$.}\label{table:s reconstruction Mtb}
\end{table}
 The cross sections of backgrounds are listed in Table~\ref{table:background recont} with the basic cuts and varying $M_{t\bar{b}}$ cuts. One can find that if a strict $M_{t\bar{b}}$ cut is adopted,  all the background effects can be neglected except for the $W$ boson process. Table~\ref{table:s reconstruction Mtb} shows the total cross sections for signal and backgrounds as well as the significance.  The number of signal events is more than 1 for $m_{W'}=6$ TeV at 14 TeV with a luminosity of $300 fb^{-1}$. The corresponding significance distribution with respect to the $W'$ mass is displayed in Fig.~{\ref{fig:72fb1314sig}}.  The upper mass limit can be up to 6.2 (6.6) TeV with a $3\sigma$ significance after we require $M_{t\bar{b}}> \frac{3}{4}m_{W^{\prime}}$ for $W'_L$ ($W'_R$) if there is no excess observed.

 Currently, the integrated luminosity is 36.1 $fb^{-1}$ reported by the ATLAS collaboration and 35.9 $fb^{-1}$ ~\cite{Aaboud:2017efa} by the CMS collaboration, with a collision energy of 13 TeV~\cite{Sirunyan:2017vkm}, so we investigate the process of Eq.~(\ref{eq:process}) at 13 TeV as well. Figure~{\ref{fig:72fb13tevsig}} displays the significance distribution with respect to the $W'$ mass with the basic cuts and a loose cut of  $M_{t\bar{b}}> \frac{2}{3}m_{W^{\prime}}$. If there is no excess observed, the $W'$ can be excluded for a mass less than 4.9 (5.6) TeV for $W'_L$ ($W'_R$) with $3\sigma$ significance.
\section{SUMMARY}\label{sec4}
 We have investigated the process $pp \to W'\slash W \to t\bar{b}\to b \bar{b} l \nu$ for the $W'$ signal via the kinematic distributions. As the signal events are characterized by $2~\text{jets}+1~\text{lepton}+~\met$, the dominant standard model backgrounds are $W^+jj$, $W^+b \bar{b}$, $W^{+}g\to t\bar{b}$, $bq \to tj$ and $t\bar{t}$. To reduce the backgrounds and improve the significance, we adopted  four schemes, i.e., the transverse momentum of jets, the invariant mass of jets, the scalar sum of the transverse momentum as well as the missing transverse energy, and the invariant mass of $t\bar{b}$ with the top quark reconstruction. By applying suitable cuts, it is possible to search for a $W'$ signal at the LHC. For example, at 14 TeV with a luminosity of 300~$fb^{-1}$, in the $H_T$ ($M_{t\bar{b}}$) scheme the $W'_R$ signal can be observed for a mass below 4.7 (6.6) TeV. These results are the consequence from the process in Eq.~(\ref{eq:process}), while the significance will be improved if the process $pp \rightarrow W'^{-} \slash W^{-} \rightarrow \bar{t}b \rightarrow b\bar{b}\ell^{-}\bar{\nu}$ is included. The aim of this paper is to investigate the possibility of searching for the  $W'$ signal in  single top production, which has the advantage of being able to scan the kinematic distribution for an excess over the standard model prediction. Once large numbers of single top quark production events have been accumulated, our methods will be helpful to search for the $W'$ signal if the new heavy resonance peak cannot be observed directly.
\section{Acknowledgements}\label{sec5}
We would like to thank the members of the particle groups at the University of Jinan and Shandong University for their helpful discussions and comments. HL and WS thank the University of Arizona for their hospitality during the writing of this paper.
This work is supported in part by the National Natural Science Foundation of China (NSFC) under grant Nos. 11325525/11635009/11775130/11447009/11305049 and  Natural Science Foundation of Shandong Province under grant Nos. ZR2017JL006/ZR2017MA002.

\clearpage
\end{CJK*}

\begin{thebibliography}{0}



\bibitem{Klein:1926tv}
  O.~Klein,
  Z.\ Phys.\  {\bf 37}, 895 (1926),
  [Surveys High Energ.\ Phys.\  {\bf 5}, 241 (1986)],
  doi:10.1007/BF01397481.

\bibitem{ArkaniHamed:1998rs}
  N.~Arkani-Hamed, S.~Dimopoulos and G.~R.~Dvali,
  Phys.\ Lett.\ B {\bf 429}, 263 (1998),
  doi:10.1016/S0370-2693(98)00466-3,
  [hep-ph/9803315].

\bibitem{Randall:1999vf}
  L.~Randall and R.~Sundrum,
  Phys.\ Rev.\ Lett.\  {\bf 83}, 4690 (1999),
  doi:10.1103/PhysRevLett.83.4690,
  [hep-th/9906064].

\bibitem{Randall:1999ee}
  L.~Randall and R.~Sundrum,
  Phys.\ Rev.\ Lett.\  {\bf 83}, 3370 (1999),
  doi:10.1103/PhysRevLett.83.3370,
  [hep-ph/9905221].

\bibitem{ArkaniHamed:2001ca}
  N.~Arkani-Hamed, A.~G.~Cohen and H.~Georgi,
  Phys.\ Rev.\ Lett.\  {\bf 86}, 4757 (2001),
  doi:10.1103/PhysRevLett.86.4757,
  [hep-th/0104005].

\bibitem{Appelquist:2000nn}
  T.~Appelquist, H.~C.~Cheng and B.~A.~Dobrescu,
  Phys.\ Rev.\ D {\bf 64}, 035002 (2001),
  doi:10.1103/PhysRevD.64.035002,
  [hep-ph/0012100].


\bibitem{Cheng:2002ab}
  H.~C.~Cheng, K.~T.~Matchev and M.~Schmaltz,
  Phys.\ Rev.\ D {\bf 66}, 056006 (2002),
  doi:10.1103/PhysRevD.66.056006,
  [hep-ph/0205314].

\bibitem{ArkaniHamed:2001nc}
  N.~Arkani-Hamed, A.~G.~Cohen and H.~Georgi,
  Phys.\ Lett.\ B {\bf 513}, 232 (2001),
  doi:10.1016/S0370-2693(01)00741-9,
  [hep-ph/0105239].

\bibitem{Han:2003wu}
  T.~Han, H.~E.~Logan, B.~McElrath and L.~T.~Wang,
  Phys.\ Rev.\ D {\bf 67}, 095004 (2003),
  doi:10.1103/PhysRevD.67.095004,
  [hep-ph/0301040].

\bibitem{Kaplan:2003uc}
  D.~E.~Kaplan and M.~Schmaltz,
  JHEP {\bf 0310}, 039 (2003),
  doi:10.1088/1126-6708/2003/10/039,
  [hep-ph/0302049].

\bibitem{Pati:1973uk}
  J.~C.~Pati and A.~Salam,
  Phys.\ Rev.\ D {\bf 8}, 1240 (1973),
  doi:10.1103/PhysRevD.8.1240.

\bibitem{Georgi:1974sy}
  H.~Georgi and S.~L.~Glashow,
  Phys.\ Rev.\ Lett.\  {\bf 32}, 438 (1974),
  doi:10.1103/PhysRevLett.32.438.


\bibitem{Fritzsch:1974nn}
  H.~Fritzsch and P.~Minkowski,
  Annals Phys.\  {\bf 93}, 193 (1975),
  doi:10.1016/0003-4916(75)90211-0.

\bibitem{Pati:1974yy}
  J.~C.~Pati and A.~Salam,
  Phys.\ Rev.\ D {\bf 10}, 275 (1974),
  Erratum: [Phys.\ Rev.\ D {\bf 11}, 703 (1975)],
  doi:10.1103/PhysRevD.10.275, 10.1103/PhysRevD.11.703.2.

\bibitem{Mohapatra:1974hk}
  R.~N.~Mohapatra and J.~C.~Pati,
  Phys.\ Rev.\ D {\bf 11}, 566 (1975),
  doi:10.1103/PhysRevD.11.566.

\bibitem{Mohapatra:1974gc}
  R.~N.~Mohapatra and J.~C.~Pati,
  Phys.\ Rev.\ D {\bf 11}, 2558 (1975),
  doi:10.1103/PhysRevD.11.2558.

\bibitem{Senjanovic:1975rk}
  G.~Senjanovic and R.~N.~Mohapatra,
  Phys.\ Rev.\ D {\bf 12}, 1502 (1975),
  doi:10.1103/PhysRevD.12.1502.

\bibitem{Mohapatra:1977mj}
  R.~N.~Mohapatra, F.~E.~Paige and D.~P.~Sidhu,
  Phys.\ Rev.\ D {\bf 17}, 2462 (1978),
  doi:10.1103/PhysRevD.17.2462.

\bibitem{Aaboud:2017efa}
  M.~Aaboud {\it et al.} [ATLAS Collaboration],
  arXiv:1706.04786 [hep-ex].

\bibitem{Khachatryan:2016jww}
  V.~Khachatryan {\it et al.} [CMS Collaboration],
  Phys.\ Lett.\ B {\bf 770}, 278 (2017),
  doi:10.1016/j.physletb.2017.04.043,
  [arXiv:1612.09274 [hep-ex]].
  
\bibitem{Sullivan:2013ina}
  Z.~Sullivan,
  arXiv:1308.3797 [hep-ph].
  
\bibitem{Duffty:2012rf}
  D.~Duffty and Z.~Sullivan,
  Phys.\ Rev.\ D {\bf 86}, 075018 (2012)
  doi:10.1103/PhysRevD.86.075018
  [arXiv:1208.4858 [hep-ph]].

\bibitem{Gopalakrishna:2010xm}
  S.~Gopalakrishna, T.~Han, I.~Lewis, Z.~G.~Si and Y.~F.~Zhou,
  Phys.\ Rev.\ D {\bf 82}, 115020 (2010),
  doi:10.1103/PhysRevD.82.115020,
  [arXiv:1008.3508 [hep-ph]].

\bibitem{Bao:2011nh}
  S.~S.~Bao, H.~L.~Li, Z.~G.~Si and Y.~F.~Zhou,
  Phys.\ Rev.\ D {\bf 83}, 115001 (2011),
  doi:10.1103/PhysRevD.83.115001,
  [arXiv:1103.1688 [hep-ph]].

\bibitem{Gong:2014qla}
  X.~Gong, H.~L.~Li, C.~F.~Qiao, Z.~G.~Si and Z.~J.~Yang,
  Phys.\ Rev.\ D {\bf 89}, no. 5, 055022 (2014),
  doi:10.1103/PhysRevD.89.055022,
  [arXiv:1403.0347 [hep-ph]].



\bibitem{Berger:2011hn}
  E.~L.~Berger, Q.~H.~Cao, C.~R.~Chen and H.~Zhang,
  Phys.\ Rev.\ D {\bf 83}, 114026 (2011)
  doi:10.1103/PhysRevD.83.114026
  [arXiv:1103.3274 [hep-ph]].


\bibitem{Berger:2011xk}
  E.~L.~Berger, Q.~H.~Cao, J.~H.~Yu and C.-P.~Yuan,
  Phys.\ Rev.\ D {\bf 84}, 095026 (2011)
  doi:10.1103/PhysRevD.84.095026
  [arXiv:1108.3613 [hep-ph]].


\bibitem{Ayazi:2010jd}
  S.~Yaser Ayazi and M.~Mohammadi Najafabadi,
  J.\ Phys.\ G {\bf 38}, 085002 (2011)
  doi:10.1088/0954-3899/38/8/085002
  [arXiv:1006.2647 [hep-ph]].


\bibitem{Kelso:2014qka}
  C.~Kelso, H.~N.~Long, R.~Martinez and F.~S.~Queiroz,
  Phys.\ Rev.\ D {\bf 90}, no. 11, 113011 (2014)
  doi:10.1103/PhysRevD.90.113011
  [arXiv:1408.6203 [hep-ph]].





\bibitem{Sirunyan:2017ukk}
  A.~M.~Sirunyan {\it et al.} [CMS Collaboration],
  JHEP {\bf 1708}, 029 (2017),
  doi:10.1007/JHEP08(2017)029,
  [arXiv:1706.04260 [hep-ex]].

\bibitem{Han:2012vk}
  T.~Han, I.~Lewis, R.~Ruiz and Z.~G.~Si,
  Phys.\ Rev.\ D {\bf 87}, no. 3, 035011 (2013),
  Erratum: [Phys.\ Rev.\ D {\bf 87}, no. 3, 039906 (2013)],
  doi:10.1103/PhysRevD.87.035011, 10.1103/PhysRevD.87.039906,
  [arXiv:1211.6447 [hep-ph]].

\bibitem{Pumplin:2002vw}
  J.~Pumplin, D.~R.~Stump, J.~Huston, H.~L.~Lai, P.~M.~Nadolsky and W.~K.~Tung,
  JHEP {\bf 0207}, 012 (2002),
  doi:10.1088/1126-6708/2002/07/012,
  [hep-ph/0201195].

\bibitem{Patrignani:2016xqp}
  C.~Patrignani {\it et al.} [Particle Data Group],
  Chin.\ Phys.\ C {\bf 40}, no. 10, 100001 (2016),
  doi:10.1088/1674-1137/40/10/100001.

\bibitem{Aad:2009wy}
  G.~Aad {\it et al.} [ATLAS Collaboration],
  arXiv:0901.0512 [hep-ex].



\bibitem{Sirunyan:2017vkm}
  A.~M.~Sirunyan {\it et al.} [CMS Collaboration],
  arXiv:1708.08539 [hep-ex].

\end{thebibliography}
  \end{document}